\documentclass[aps,pre,floatfix,nofootinbib,showpacs,twocolumn]{revtex4}
\usepackage{graphicx}
\usepackage{amsmath}
\usepackage{amssymb}

\newcommand{\BEQ}{\begin{eqnarray}}
\newcommand{\EEQ}{\end{eqnarray}}
\newcommand{\BEA}{\begin{eqnarray}}
\newcommand{\EEA}{\end{eqnarray}}

\renewcommand{\d}{{\rm d}}

\newcommand{\ep}{\varepsilon}

\newcommand{\eps}{\varepsilon}

\newcommand{\tr}{{\rm tr}}

\newcommand{\W}{{\mathfrak W}}

\newcommand{\Ce}{{\mathfrak e}}

\renewcommand{\S}{{\rm S}}
\newcommand{\R}{{\rm R}}

\renewcommand{\u}{u_{\rm m}} 
\renewcommand{\v}{v_{\rm m}}

\newcommand{\diag}{{\rm diag}}
\begin{document} 
\draft
\title
{
Work extremum principle: Structure and function of quantum heat engines
}
\date{\today}
\author{Armen E. Allahverdyan}
\affiliation{Yerevan Physics Institute,
Alikhanian Brothers Street 2, Yerevan 375036, Armenia}
\author{Ramandeep S. Johal}
\affiliation{Post-Graduate Department of Physics,
Lyallpur Khalsa College, Jalandhar 144001, Punjab, India}
\author{Guenter Mahler}
\affiliation{ Institute of Theoretical Physics I, University of Stuttgart, 
Pfaffenwaldring 57, 70550 Stuttgart, Germany}

\begin{abstract}

We consider a class of quantum heat engines consisting of two subsystems
interacting via a unitary transformation and coupled to two separate
baths at different temperatures $T_h>T_c$. The purpose of the engine is
to extract work due to the temperature difference. Its dynamics is not
restricted to the near equilibrium regime. The engine structure is
determined by maximizing the extracted work under various constraints.
When this maximization is carried out at finite power, the engine
dynamics is described by well-defined temperatures and satisfies the
local version of the second law. In addition, its efficiency is bounded
from below by the Curzon-Ahlborn value $1-\sqrt{T_c/T_h}$ and from above
by the Carnot value $1-(T_c/T_h)$. The latter is reached|at finite
power|for a macroscopic engine, while the former is achieved in the
equilibrium limit $T_h\to T_c$. When the work is maximized at a zero
power, even a small (few-level) engine extracts work right at the Carnot
efficiency.

\end{abstract}



\maketitle

\section{Introduction.}

Heat engines are natural or artificial devices, the goal of which is
to extract work (high-graded energy) from non-equilibrium sources of
heat (low-graded energy) \cite{landau,callen}.  There are {\it three}
basic characteristics of the engine operation: {\it i)} the work
extracted per cycle; {\it ii)} the efficiency with
which the input heat is converted into work; {\it iii)} the power of
work-extraction, i.e.  the work-extracted per cycle divided over the
cycle duration. 

Heat engines may evolve in time toward
states that provide (constrained) optimization of their functional
characteristics. Here are three examples from largely different fields:

{\bf 1.} The modern artificial engines are much more powerful and
efficient than those which started the Industrial Revolution.  The
cause of this improvement are engineering efforts driven by
our desire to get more high-graded energy at a lower cost. 

{\bf 2.} A similar trend is seen in biology and ecology. Driven by
evolution, higher organisms and more developed ecosystems have more
refined and more optimal means of extracting energy from their respective
environment \cite{zotin}.  This observation led to several quantitative
formulations, which were verified experimentally \cite{zotin}. 

{\bf 3.} The earth atmosphere can approximately be regarded as a 
huge heat-engine
operating between two thermal baths (cold bath of the earth surface and
eventually the hot bath of the sun) and producing as output large-scale turbulent
motion of air and vapor \cite{geo}. As verified by observation
\cite{geo}, this engine is also tuned to extract the maximal work
\cite{geo}, although the precise mechanism of this tuning is unclear yet. 

Needless to stress that in all above situation the optimization of heat
engines proceeds in the presence of constraints that 
determine the very path of the engine evolution.  This fact is especially
important in the above bio-ecological perspective.

The purpose of this paper is to understand the structure of quantum heat
engines emerging from the maximization of work (produced per cycle)
under specific constraints.  The study of quantum engines started in
60's \cite{60}, when it was realized that many models of lasers and
masers are in fact quantum heat engines. A good review of this early
activity is given in \cite{review}.  Nowadays the physics of quantum
heat engines is a rich field\cite{kosloff,geva,var,nori,nori_c,beretta,boukobza,linke,linke_talk,ki,scully,1,2,alicki}
related to other branches of modern quantum theory; see \cite{muru} for
a recent review of engines in the context of quantum information theory.
We see two basic reasons for studying quantum heat engines: {\it i)}
understanding of how thermodynamics emerges from the quantum mechanics;
{\it ii)} clarifying principal possibilities of nano-scale devices. 

Our analysis starts from quantum mechanics and
does not rely on the validity of thermodynamical concepts normally
invoked in studying heat engines \cite{landau,callen}.  (Relying on
quantum mechanics does not preclude us from considering macroscopic
systems.) We allow the intermediate states of the engine to be arbitrary
far from equilibrium.  We shall however see that local thermodynamical
concepts|such as the existence of local temperatures in the intermediate
stages of the engine functioning, or the validity of local formulations
of the second law|emerge as a result of maximizing the produced work. 

\begin{figure}
\includegraphics[width=7cm]{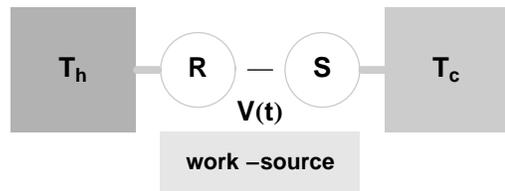}
\caption{A
schematic representation of the considered quantum engine. Quantum
systems $\R$ and $\S$ interact with thermal baths at temperatures $T_h$
and $T_c$, respectively. The mutual interaction between them is governed by
the potential $V(t)$ that also serves to deliver work to an external
source. The mutual interaction is switched on for a short time only.
Once it is switched off, the systems $\S$ and $\R$ do relax to
equilibrium under influence of the respective thermal baths. Similar constructions
of quantum heat engines appeared in \cite{kosloff,nori,jmod}.
}
\hfill
\label{fig_0}
\end{figure}

Our model will consist of two quantum systems $\R$ and $\S$ interacting
with thermal baths at temperatures $T_h$ and $T_c$, respectively
($T_h>T_c$); see Fig.~\ref{fig_0}. The number of energy
levels for $\R$ and $\S$ is finite, but it can be made very large going
to the macroscopic limit. $\R$ and $\S$ interact with an external
work-source producing work and then relax back to equilibrium under
influence of the baths. For given temperatures (and given number of
subsystem energy levels) work maximization can be introduced on three
different levels: (1.) One optimizes the extracted work over the
interaction of $\R$ and $\S$ with the external source of work (2.) One
maximizes the work, in addition, over the spectral structure of $\R$ and
$\S$.  (3.) Finally, one optimizes the extracted work also over the
interaction of $\R$ and $\S$ with their respective thermal baths.  In
the following we will investigate these strategies in full detail. 

Our main results are that on the second level of optimization the
efficiency at the maximal work is always bounded from below by the
Curzon-Ahlborn value $1-\sqrt{T_c/T_h}$. This is in addition to the
upper bound given by the Carnot value $1-(T_c/T_h)$, which for the
present setup holds generally, i.e., without any optimization.  For any
number of energy levels of $\R$ and $\S$ the Curzon-Ahlborn value is
reached in the equilibrium limit $T_h\to T_c$. When the number of levels
for $\R$ and $\S$ become very large (macroscopic limit), the efficiency
reaches the Carnot value $1-(T_c/T_h)$.  In contrast to the standard
thermodynamic realization of the Carnot efficiency, in our situation
this efficiency is reached at a finite power.  The third level of
optimization is related to making the power very small.
Now even a finite (two)-level systems $\R$ and $\S$ may be employed for
extraction of a finite [per cycle] amount of work right at the Carnot
efficiency. 

This paper is organized as follows. Section \ref{model} describes the
studied model for a quantum heat engine.  Here we also recall the
derivation of the Carnot bound for the efficiency and address the
problem of the power of work. In section \ref{III} we discuss the
optimization of work over the interaction with the sources.  In section
\ref{IV} we report the results emerged from maximizing the work over the
spectral structure of the involved quantum systems. In particular, we
show that the Curzon-Ahlborn value is a lower bound for the efficiency,
and we describe how one can employ a macroscopic engine for
work-extraction at a finite power and with the Carnot efficiency.  In
section \ref{V} we sacrify the finite-power condition, and show that a
finite-level quantum engine can extract work right at the Carnot
efficiency.  The last section presents our conclusions, discusses some
open questions and compares our findings with the results obtained
previously.  Several technical issues are discussed in appendices.

\section{The model}
\label{model}

Our engine consists of two quantum systems $\R$ and $\S$ which
individually and in parallel couple to two different thermal baths and
interact with a source of work; see Fig.~\ref{fig_0}. In the first step
$\R$ and $\S$ undergo a work extraction process driven by the source.
In the second step $\R$ and $\S$ relax back to their initial states
under influence of the baths.  Thus thermodynamic machine cycle is
performed. 

\subsection{Initialization}

Consider the two finite quantum systems $\R$ and $\S$ with Hamiltonians $H_\R$
and $H_\S$, respectively. R (S) has $n$ ($m$) energy levels. 
In the scenarios studied here almost all operators involved will be 
diagonal in the energy representation. Thus we write
\BEA
H_\R=\diag [\eps_n,...,\eps_1], \quad \eps_n\geq ...\geq \eps_1, \\
H_\S=\diag [\mu_m,...,\mu_1], \quad \mu_m\geq ...\geq \mu_1,
\label{01}
\EEA
where $\diag[a,..,b]$ is a diagonal matrix with entries $(a,...,b)$.
We will take
\BEA
\eps_1=\mu_1=0, 
\EEA
that is the energies of both systems are measured from zero. 
It will be seen below that this choice does not imply any loss of generality.
Thus, $\eps_k$ and $\mu_l$ are the energy level spacings. 
The total Hamiltonian is
\BEA
\label{ham}\label{lala1}
&&H=H_\R\otimes 1+1\otimes H_\S=\sum_{\alpha=1}^{nm}E_\alpha |E_\alpha\rangle\langle E_\alpha|,
\\
&&=\diag[
\eps_n+\mu_n,\eps_n+\mu_{n-1},...,\eps_1+\mu_2,\eps_1+\mu_1],
\label{ham1}
\EEA
where $\{E_\alpha\}_{\alpha=1}^{nm}$ and
$\{|E_\alpha\rangle\}_{\alpha=1}^{nm}$ are, respectively, eigenvalues and 
eigenvectors of $H_R\otimes 1+1\otimes H_S$.

Subsystem $\R$ ($\S$)is now assumed to couple to 
a thermal bath at temperature $T_h$ ($T_c$); see Fig.~\ref{fig_0}. 
We shall assume:
\BEA
T_h>T_c.
\label{non}
\EEA

At this stage there is no mutual coupling between R and S, and each
bath drives its corresponding subsystem to an equilibrium state specified 
by the Gibbss density matrices
\BEA
\label{1}
\rho=\frac{e^{-\beta_h H_\R}}{\tr\, e^{-\beta_h H_\R}}, \quad
\sigma=\frac{e^{-\beta_c H_\S}}{\tr\, e^{-\beta_c H_\S}}, 
\EEA
where $T_h=1/\beta_h$ and $T_c=1/\beta_c$. Alternatively,
\BEA
\label{kro1}
&&\rho=\diag [r_n,...,r_1], \quad r_n\leq ...\leq r_1, \\
\label{kro2}
&&\sigma=\diag [s_m,...,s_1], \quad s_m\leq ...\leq s_1,\\
\label{kuban1}
&&\eps_k=T_h\ln\frac{r_1}{r_k}, \quad k=1,...,n,\\
&&\mu_l=T_c\ln\frac{s_1}{s_l}, \quad l=1,...,m.
\label{kuban2}
\EEA
The overall initial density matrix is 
\BEA
\label{omega_in}
\label{lala2}
\Omega_{\rm in}&=&\rho\otimes\sigma=\sum_{\alpha=1}^{nm}P_\alpha |P_\alpha\rangle\langle P_\alpha| \\
&=&\diag[
r_ns_n, r_ns_{n-1},r_ns_{n-2},..., r_1s_1],
\label{omega_in1}
\EEA
and the average initial energy 
\BEA
\label{art0}
\tr [\, H\, \Omega_{\rm in} \,]
=\sum_{\alpha=1}^{nm}P_\alpha E_\alpha.
\EEA

\subsection{Two step process}

\subsubsection{First step: Unitary transformation}
\label{first_step}

Now $\R+\S$ are taken to interact via a time-dependent potential $V(t)$ so that 
their Hamiltonian reads
\BEA
\label{ham_ham}
{\cal H}(t)=H_\R\otimes 1+1\otimes H_\S+V(t),\\
\label{cyclop}
V(t)=0\quad {\rm for} \quad t<0 \quad {\rm and~~~ for} \quad t>\tau.  
\EEA
This means that the initial and final Hamiltonian (i.e. before and after this
first step) are both given by (\ref{ham}).
The potential $V(t)$ in (\ref{ham_ham}) is assumed to be sufficiently strong 
[and to act
in a sufficiently short time] so that the influence of the thermal baths
between the 
times $0$ and $\tau$ can be neglected. 
We require this step to be thermally
isolated, since, otherwise, for the couplings to the baths being on, one
would get a direct heat exchange between the baths, a factor that should
decrease the overall efficiency of the work-extraction. 
Thus the dynamics of $\R+\S$ is unitary for
$0\leq t\leq \tau$:
\BEA
\Omega_{\rm fin}={\cal U}\, \Omega_{\rm in}\, {\cal U}^\dagger, \quad
{\cal U}={\cal T}\exp\left[-i\int_0^\tau \d \bar{t}\,{\cal H}(\bar{t}) 
\right],
\label{klein}
\EEA
where $\Omega_{\rm in}$ is the initial density matrix defined in
(\ref{omega_in}), $\Omega_{\rm fin}$ is the final density matrix, ${\cal
U}$ is the unitary evolution operator, and  ${\cal T}\exp$ is the
chronologically ordered exponent.  
The work extracted in this step reads
\BEA
W_{nm}(T_h,T_c,\eps,\mu,V)=\tr [\, H\, (\Omega_{\rm in}-\Omega_{\rm fin}) \,], 
\label{work}
\EEA
This work $W$ depends on a set of parameters: The number of energy
levels $n$ and $m$ for $\R$ and $\S$, respectively, the spectra $\eps$
and $\mu$ according to (\ref{01}), and the potential $V(t)$. In the
following we do not make explicit the dependence of $W$ on the
temperatures $T_h$ and $T_c$, which are taken to be fixed conditions in
all cases.

\subsubsection{Second step: Relaxation}
\label{efo}

After $V(t)$ has been switched off, which means that the final
Hamiltonian is again given by (\ref{ham}), R and S (after some
relaxation time) return to the initial states (\ref{1}, \ref{omega_in}),
under influence of the two thermal baths as introduced before.  This
completes the two-step cycle; now the same heat to work transformation
can be repeated. 

During the relaxation the thermal baths at temperatures $T_h$ and $T_c$
get, respectively, the amounts of heat 
\BEA
\label{am3}
Q_h=\tr (H_R\,\rho)- \tr (H_R\otimes 1\,\Omega_{\rm fin}), \\
Q_c=\tr (H_S\,\sigma)- \tr (H_S\otimes 1\,\Omega_{\rm fin}).
\label{am333}
\EEA
The work (\ref{work}) is expressed as $W_{nm}(\eps,\mu,V)=Q_h+Q_c$.

Following to \cite{partovi,jmod} we determine how $Q_h$ and $Q_c$ relate to the
temperatures. Recall that the relative entropy between two density
matrices $\rho_1$ and $\rho_2$ is defined as \cite{delgado}
\BEA
\label{relative_entropy}
S[\rho_1||\rho_2]={\rm
tr}[\rho_1\ln \rho_1-\rho_1\ln\rho_2].  
\EEA
This non-negative quantity 
reflects the difference between $\rho_1$ and $\rho_2$: $S[\rho_1||\rho_2]=0$
if and only if $\rho_1=\rho_2$ \cite{delgado}. 
Due to the unitarity of the work-extracting
process (\ref{klein}), ${\rm tr}[\Omega_{\rm in}\ln \Omega_{\rm in}]={\rm
tr}[\Omega_{\rm fin}\ln \Omega_{\rm fin}]$, where $\Omega_{\rm in}$ and
$\Omega_{\rm fin}$ are defined by (\ref{lala2}) and (\ref{klein}). We get
\BEA
\label{oslo}
-S[\Omega_{\rm fin}|| \Omega_{\rm in}]=
\frac{Q_h}{T_h}+\frac{Q_c}{T_c}
\leq 0.
\EEA
If a non-negative amount of work is extracted, 
$W_{nm}(\eps,\mu,V)=Q_h+Q_c\geq 0$, Eq.~(\ref{oslo}) 
implies $W_{nm}(\eps,\mu,V)\leq (1-\frac{T_c}{T_h})Q_h$. Together with
$T_h>T_c$ this means that the heat flows from the higher temperature to the lower one:
\BEA
Q_h\geq 0\quad {\rm and} \quad Q_c\leq 0. 
\label{g_lobo}
\EEA
For the efficiency of the work-extraction we get that it
is always bound from above by the Carnot value
\BEA
\label{ef}
\eta_{nm}(\eps,\mu,V)&\equiv&\frac{W_{nm}(\eps,\mu,V)}{Q_h}\\
\label{carnot}
&\leq& \eta_{\rm Carnot}\equiv 1-\frac{T_c}{T_h}.
\EEA
Eqs.(\ref{g_lobo}--\ref{carnot}) are, of course, well known in the ordinary thermodynamics.
The advantage of the above derivation is that their validity is
confirmed out of the local equilibrium \cite{jmod} and without requiring
a weak system-bath coupling. The latter assumption for deriving
the Carnot bound is employed in Ref.~\cite{alicki_jpa}.

\subsection{Power of work extraction and cycle time}
\label{power}

As discussed above, the engine operation consists of two steps: work
extraction during a reversible unitary transformation, which (in
principle) can be implemented in an arbitrary short time, and the heat
gathering step, when the subsystems R and S relax back to their initial
states under influence of the baths. This relaxation is in general
irreversible, since, as will be seen below, at the moment when the
second stage starts, the state of $\R$ ($\S$) as given by
(\ref{omega_fin}) is not at local equilibrium with the hot (cold)
bath. 

If the couplings of R and S to their baths were weak, the second step
would take a very long time and thus the power|the work extracted per
cycle divided over the cycle time|would tend to zero. 

However, the second step can also be implemented in a finite time: We
may then assume that the couplings to the bath are finite, or even
large, so that the relaxation to the states (\ref{1}) does not take very
long, but the work-extraction stage is still, to a good approximation,
thermally isolated. In Appendix \ref{app_collisions} we describe a realistic
relaxation mechanism that achieves a finite-time relaxation, but still 
allows a thermally-isolated work-extracting transformation, as described in 
section \ref{first_step}.

Thus provided we extract a finite amount of work during the first stage,
we can get a finite power of work-extraction if
the couplings of R and S to their baths are not weak.

\section{Maximizing the work over the potential $V(t)$}
\label{III}

We first optimize the work $W_{nm}(\eps,\mu,V)$ in (\ref{work}) over all unitary
evolutions (\ref{klein}). This is equivalent to maximizing $W_{nm}(\eps,\mu,V)$
over all potentials $V(t)$, defined in (\ref{ham_ham}, \ref{cyclop}):
\BEA
\label{ttu}
W_{nm}(\eps,\mu)&\equiv& {\rm max}_{\,V\,}[\,W_{nm}(\eps,\mu,V)\,]
\EEA

This is a known problem in thermodynamics \cite{landau,callen}.
However, the standard textbook answer to this problem applies only to
macroscopic, weakly-non-equilibrium systems S and R.  The general
solution to the maximal work-extraction problem for a given initial
state $\Omega_{\rm in}$ was given in \cite{abn}.  For the considered
situation (\ref{1}--\ref{omega_in}) this solution amounts to the
following \cite{abn}. 
Once the initial state (and thus the initial energy) is fixed,
for maximizing the work (\ref{work})
one has to minimize the final energy (cf. (\ref{omega_in}--\ref{art0})):
\BEA
\label{art}
\tr [\, H\, \Omega_{\rm fin} \,]
=\sum_{\alpha=1}^{nm}\sum_{\beta=1}^{nm}
C_{\alpha\beta}P_\beta E_\alpha,
\EEA
where 
\BEA
\label{bron}
C_{\alpha\beta}=\langle E_\alpha|{\cal U}|E_\beta\rangle 
\langle E_\beta|{\cal U}^\dagger|E_\alpha\rangle, 
\EEA

Note that
$C_{\alpha\beta}$ is a double-stochastic matrix, i.e., {\it i)}
$C_{\alpha\beta}\geq 0$, {\it ii)} $\sum_{\alpha=1}^n C_{\alpha\beta}=1$
and $\sum_{\beta=1}^n C_{\alpha\beta}=1$. Conversely, any matrix that
satisfies the features {\it i)} and {\it ii)} can be presented as in
(\ref{bron}) for some unitary operator ${\cal U}$ \cite{ando}.  A
particular case of double-stochastic matrices are permutation matrices,
which on each row and on each column contain one element equal to $1$,
while other elements are equal to zero. For this simplest case, which
will be most relevant  for our purpose, the relation with
unitaries is especially clear, since the transpose $\Pi^T$ of a permutation matrix
$\Pi$ coincides with its inverse $\Pi^{-1}$. Thus, $\Pi$ is already
orthogonal (unitary).

Now it is obvious \cite{ando} which double-stochastic matrix $C_{\alpha\beta}$
minimizes the sum in the RHS of (\ref{art}). This is a permutation
matrix that enforces the largest element in $E$ to appear in (\ref{art})
paired with the smallest element in $P$. The one but the largest in $E$
gets the one but the smallest element in $P$, and so on.  Finally, the
smallest element in $P$ is paired with the largest element in $E$ \cite{ando}.

To write this down, let $\Pi_H$ 
be the permutation that orders the sequence $E$
in (\ref{ham1}) in the non-increasing order
\BEA
\label{lala3}
(\Pi_H E)_{nm}
\geq (\Pi_H E)_{nm-1}\geq ...\geq (\Pi_H E)_1,
\EEA
where $(\Pi_H E)_{nm}$ is the largest element of $\Pi_H E$ which is
obviously equal to $\eps_n+\mu_n$ due to (\ref{kro1}), $(\Pi_H
E)_{nm-1}$ is the one but the largest element of $E$ which in general is
not anymore equal to $\eps_{n}+\mu_{n-1}$, and {\it etc.}

In the same fashion, let $\Pi_\Omega$ be the permutation that orders 
in the non-decreasing order the sequence $P$ in (\ref{omega_in1}):
\BEA
\label{lala4}
(\Pi_\Omega P)_{nm}
\leq (\Pi_\Omega P)_{nm-1}\leq ...\leq (\Pi_\Omega P)_1.
\EEA
Note that $(\Pi_\Omega P)_{nm}=r_ns_m$ and $(\Pi_\Omega P)_{11}=r_1s_1$ due to (\ref{kro2}).
Then the minimal final energy can be written as $\sum_{\alpha=1}^{nm} 
(\Pi_\Omega P)_{\alpha} (\Pi_H E)_{\alpha}=\sum_{\alpha=1}^{nm} 
(\Pi_H^{-1}\Pi_\Omega P)_{\alpha} E_{\alpha}$, where $\Pi_H^{-1}$ is 
the inverse permutation to $\Pi_H$:
$\Pi_H^{-1}\Pi_H=1$. The relation between $\Pi_H^{-1}\Pi_\Omega P$ and $E$
is that if for any two indices $E_\alpha\geq E_\beta$, then
$(\Pi_H^{-1}\Pi_\Omega P)_\alpha\leq (\Pi_H^{-1}\Pi_\Omega P)_\beta$.

Thus the maximum of the work $W$ over all interactions $V(t)$ in
(\ref{ham_ham}) is equal to [recall (\ref{art0})]
\BEA
W_{nm}(\eps, \mu )=\sum_{\alpha=1}^{nm} P_{\alpha} E_{\alpha}
-\sum_{\alpha=1}^{nm} 
(\Pi_H^{-1}\Pi_\Omega P)_{\alpha} E_{\alpha}.
\label{am2}
\EEA
The maximum (\ref{am2}) is achieved for the final state
\BEA
\label{omega_fin}
\Omega_{\rm fin} 
= \diag[\Pi_H^{-1}\Pi_\Omega P],
\EEA
which obviously has the same (but differently ordered) eigenvalues as
$\Omega_{\rm in}$, since eigenvalues (but not their order) are
invariants of any unitary transformation.  Thus $\tr (\,\diag[\Pi_H
E]\,\,\diag[\Pi_\Omega P]\,)=\sum_{\alpha=1}^{nm} (\Pi_H^{-1}\Pi_\Omega
P)_{\alpha} E_{\alpha}$ is the lowest possible energy that can be
achieved by permuting the eigenvalues of $\Omega_{\rm in}$ \cite{abn}. 

Introducing the separate final states of R and S
\BEA
\label{marginal_final}
\rho_{\rm fin}={\rm tr}_{\rm S} \Omega_{\rm fin},\quad
\sigma_{\rm fin}={\rm tr}_{\rm R} \Omega_{\rm fin},
\EEA
and using (\ref{cyclop}) one can write 
\BEA
W_{nm}(\eps,\mu)={\rm tr}\left[ (\rho-\rho_{\rm fin})H_R\right]+
  {\rm tr}\left[ (\sigma-\sigma_{\rm fin})H_S\right].
\label{marginal_work}
\EEA

Note that $\rho_{\rm fin}$ and $\sigma_{\rm fin}$
commute with the respective Hamiltonians:
\BEA
\label{kasablanka}
[\rho_{\rm fin}, H_R]=0,\qquad [\sigma_{\rm fin}, H_S]=0,
\EEA
and that they provide a larger probability for a smaller energy, i.e.,
the analog of (\ref{kro1}, \ref{kro2}) is valid for $\rho_{\rm fin}$ and
$\sigma_{\rm fin}$. Would not these both properties have to hold, one
could extract more work from R and S. Note as well that in general
$\rho_{\rm fin}$ and $\sigma_{\rm fin}$ do not have the Gibbsian form,
i.e., they are not described by definite temperatures. 

The general form of the maximal work extracting interaction $V(t)$
(see (\ref{ham_ham}) for the definition) is given in \cite{abn}.

\section{Maximizing the work over the subsystem spectra}
\label{IV}

\subsection{Optimal spectral form}
\label{la_Habana}

We saw how to maximize the work over all interactions $V(t)$.  We
now maximize the work $W_{nm}(\eps,\mu)$ also over all energy level
spacings $\{\eps_k\}_{k=1}^n$ and $\{\mu_l\}_{l=1}^m$, i.e. over the
initial state (\ref{omega_in}).  The obtained value of work $W$ will
still be a function of $T_h$, $T_c$, $n$ and $m$. 

So far we have not been able to carry out analytically the optimization of
$W$ over all possible energy spacings. Thus we have to settle for a numerical 
optimization of
$W$ employing the standard optimization packages of Mathematica 5
\footnote{\label{comma}More specifically, we used three packages based respectively on
genetic algorithms, random-gradient search and stimulated annealing, to
ensure that we get the correct results.}. 
The result that emerged for 
\BEA
n=m
\EEA 
has the following simple form: The maximal work  
\BEA
\label{bobo}
W_{nm}\equiv {\rm max}_{\,\eps,\,\mu\,}[\,  W_{nm}(\eps, \mu)\,]
\EEA
is attained for
\BEA
\label{go1}
\eps_n=\eps_{n-1}=...=\eps_2\, >\, \eps_1=0, \\
\mu_n=\mu_{n-1}=...=\mu_2\, >\, \mu_1=0,
\label{go2}
\EEA
that is when both upper levels $\eps_2$ and $\mu_2$ are $(n-1)$ fold
degenerate \footnote{Appendix \ref{lion} discusses in which sense the system
with spectrum (\ref{go1}) is equivalent to a two-level system}.
This result has been checked numerically for $n=3,4,5$ and we
expect it to hold for an arbitrary $n=m$. The intuition behind (\ref{go1},
\ref{go2}) is that the energy is concentrated|as much as the thermal
equilibrium allows|at higher energy levels creating a sort of
instability that facilitates the subsequent work-extraction. 

For the occupation numbers one gets from (\ref{go1}, \ref{go2}):
\BEA
\label{go3}
r_n=r_{n-1}=...=r_2\, <\, r_1, \\
s_n=s_{n-1}=...=s_2\, <\, s_1.
\label{go4}
\EEA

In the context of (\ref{go1}--\ref{go4}) we note that any state with
occupation (\ref{go3}) and energies (\ref{go1}) admits a well defined
temperature. While we shall elaborate on this point in sections
\ref{kapo1} and \ref{kapo2} below, it is important to stress already now
that the existence of final temperatures came out as the result of
maximization: in general the final state (\ref{marginal_final}) do not
admit any well-defined temperature. 

Once the fact of (\ref{go1}, \ref{go2}) is recognized, it is not
difficult to get the explicit expressions for the maximal work and the
corresponding efficiency at maximal work. The maximization of work
will be split into two parts. First we shall explore several
consequences of conditions (\ref{go1}, \ref{go2}) treating $\eps_n$ and
$\mu_n$ as parameters. Then in section \ref{maxw} we shall optimize the work
over $\eps_n$ and $\mu_n$.

Assuming (\ref{go1}, \ref{go2}) we get from (\ref{lala1}, \ref{go1}, \ref{go2}):
\BEA
E=[\underbrace{\underbrace{\eps_2+\mu_2}_{n-1\,\,{\rm times}},\,\, \eps_2}_{n-1\,\,{\rm times}}, \,\,\,
   \underbrace{\mu_2}_{n-1\,{\rm times}},\,\,\, 0]
\EEA

Analogously we get from (\ref{lala2}, \ref{go3}, \ref{go4}):
\BEA
P=[\underbrace{\underbrace{r_2\,s_2}_{n-1\,\,{\rm times}},\,\, \underline{r_2\,s_1}}_{n-1\,\,{\rm times}}, \,\,\,
   \underbrace{\underline{r_1\,s_2}}_{n-1\,{\rm times}},\,\,\, r_1\,s_1]
\label{fr}
\EEA

When comparing $E$ with $P$, we see that 
the action of $\Pi_E^{-1}\Pi_\Omega$, as defined in (\ref{lala3}, \ref{lala4}), 
amounts to interchanging the underlined elements of $P$ in (\ref{fr}):
\BEA
\Pi_H^{-1}\Pi_\Omega P=
[\underbrace{\underbrace{r_2\,s_2}_{n-1\,\,{\rm times}},\,\, r_1\,s_2}_{n-1\,\,{\rm times}}, \,\,\,
   \underbrace{r_2\,s_1}_{n-1\,{\rm times}},\,\,\, r_1\,s_1]
\label{lena}
\EEA

One gets from (\ref{am2}):
\BEA
\label{tarakan}
W_{nn}(\eps_2,\mu_2)=(n-1)(\eps_2-\mu_2)(r_2s_1-r_1s_2).
\EEA
This equation can be re-written in a more convenient form by introducing 
new variables (Boltzmann weights):
\BEA
\label{bzez1}
u\equiv e^{-\beta_h \eps_2}, \qquad
v\equiv e^{-\beta_c \mu_2},
\EEA
so that
\begin{gather}
\label{bzez2}
r_2=\frac{u}{1+(n-1)u}, \qquad r_1=\frac{1}{1+(n-1)u},\\
\label{bzez3}
s_2=\frac{v}{1+(n-1)v}, \qquad s_1=\frac{1}{1+(n-1)v},\\
\label{bzez4}
\eps_2= T_h\ln\frac{1}{u},\qquad \mu_2= T_c\ln\frac{1}{v}.
\end{gather}
Eq.~(\ref{tarakan}) thus reads
\begin{gather}
\label{kora1}
W_{nn}(u,v)
=\frac{(n-1)[u-v]\left[T_h\ln\frac{1}{u}-T_c\ln\frac{1}{v}
\right]}{(1+(n-1)u)(1+(n-1)v)}.
\end{gather}
Recalling (\ref{non}), 
for the positivity of $W_{nn}(u,v)$ in (\ref{kora1}) it is necessary to have
\footnote{ The inverse conditions $u<v$ and $T_h\ln \frac{1}{u}<T_c\ln \frac{1}{v}$
are not compatible with each other due to condition (\ref{non}).} 
\BEA
\label{mushi}
u>v \qquad {\rm and} \qquad T_h\ln \frac{1}{u}>T_c\ln \frac{1}{v}, 
\EEA
which together with (\ref{non}, \ref{bzez1}, \ref{go1}, \ref{go2}) leads to
\BEA
\label{shushi}
1<\frac{\eps_2}{\mu_2}<\frac{T_h}{T_c}.
\EEA

For the efficiency of work-extraction we get from (\ref{am3}, \ref{ef}) and 
(\ref{mushi}):
\footnote{Note that from (\ref{bzez1}) and (\ref{kora2})
that for a fixed $\eps_2$ and $\mu_2$ we get the well known Otto-cycle
result $\eta = 1 - \frac{\mu_2}{\epsilon_2}$, which is constrained by
(\ref{shushi}), but otherwise does not depend on temperatures. The Otto
cycle is realized via two isothermal and two adiabatically [slow]
processes; see, e.g., \cite{nori,ki}. Thus in this scenario the power of work
is very small.  In contrast, as we stressed already, in our situation
the efficiency (\ref{kora2}) is obtained with a finite power.  }
\BEA
\label{kora2}
\eta_{nn}(u,v)=1-\frac{{\rm min}\left[ T_h\ln\frac{1}{ u}, T_c\ln\frac{1}{ v}  \right]}
{{\rm max}\left[ T_h\ln\frac{1}{ u}, T_c\ln\frac{1}{ v}  \right]}
=1-\frac{T_c\ln\frac{1}{ v} }
{T_h\ln\frac{1}{ u}}.~
\EEA

\subsubsection{Carnot limit.}
\label{carnot_finite}

Comparing (\ref{kora1}) with (\ref{kora2}) we see that for a finite $n$
the efficiency tending to the Carnot value $1-T_c/T_h$ means that the
work extracted per cycle goes to zero [recall that the cycle duration is
finite]. In section \ref{macro_macro} we show that for $n\to \infty$
this conclusion changes, and a finite amount of work can be extracted
with the efficiency [almost] equal to the Carnot value. 

Note that the work turning to zero at the Ganot limit is typical for
quantum realizations of the Otto cycle \cite{ki,nori}. However, there
also the cycle duration is very long.

\subsection{Structure of the target state.}

\subsubsection{Optimal work-extracting transformation.}

It is seen from (\ref{omega_fin}, \ref{fr}, \ref{lena}) that
under conditions (\ref{go1}, \ref{go2}) the 
optimal work extracting evolution amounts
to exchanging the states of S and R:
\BEA
\Omega_{\rm in}=\rho\otimes \sigma, 
\qquad
\Omega_{\rm fin}
=\sigma\otimes \rho.  
\label{duko}
\EEA
The meaning of this factorization is discussed
after (\ref{roma}).

The optimal work-extraction unitary operation is the so called
SWAP, well known|especially for the particular case $n=2$|as one of the
basic gates of quantum computation \cite{delgado}. Although SWAP is
normally realized via composition of several more elementary unitary
operations \cite{delgado}, its direct implementation for
atomic few-level systems was argued to be feasible \cite{jauslin}. 

\subsubsection{Temperatures after work-extraction.}
\label{kapo1}

Recall with (\ref{non}) that $\Omega_{\rm in}$ is a non-equilibrium
state.  The final density matrix $\Omega_{\rm fin}$ is also
non-equilibrium, but it is expected to be closer to equilibrium than
$\Omega_{\rm in}$. 

Let us see in which sense this expectation is going to be correct.
Recalling (\ref{kuban1}, \ref{kuban2}) 
and (\ref{go1}, \ref{go2}), the
final temperatures $T_h'$ and $T_c'$ of R and S, respectively, is
deduced from (\ref{duko})
\begin{gather}
e^{-\beta_h' \eps_2}=\frac{s_2}{s_1}=v, \quad 
e^{-\beta_c' \mu_2}=\frac{r_2}{r_1}=u,\\
T_h'=T_h\,\frac{\ln \frac{1}{u}}{\ln \frac{1}{v}}, 
\quad
T_c'=T_c\,\frac{\ln \frac{1}{v}}{\ln \frac{1}{u}}.
\label{brams} 
\end{gather}
With help of (\ref{kora2}) and of the Carnot bound (\ref{carnot}) we get
\BEA
\label{mozart}
T_h>T_h' , \qquad T_c<T_c'.
\EEA
Thus the initially hotter system R 
cools down, while initially cooler
S heats up. Recall that the physical meaning of the Carnot bound is closely tied
to the second law, i.e., to the impossibility of transferring heat completely 
into work. Eqs.~(\ref{mozart}) provide a somewhat different perspective on the Carnot bound related
to the zeroth law.

Another inequality is derived via 
(\ref{brams}, \ref{shushi}):
\BEA
\frac{T_h'}{T_c'}
<\frac{\eps_2}{\mu_2}< \frac{T_h}{T_c} \frac{T_h'}{T_c'}. 
\label{globus}
\EEA
This means that the analog of the condition (\ref{shushi}) is not
satisfied for the final temperatures ${T_c'}$ and ${T_h'}$, so that 
one cannot, again,  employ the final state in (\ref{duko}) for work-extraction via a
thermally isolated process. 

\subsubsection{Relation with the Curzon-Ahlborn efficiency.}
\label{kapo2}

If $\R$ initially
having a higher temperature than $\S$ still has this property
after the work extraction, i.e., if
\BEA
\label{hub}
T_h'>T_c',
\EEA
we get, employing (\ref{brams}) that (\ref{hub}) is equivalent to 
\BEA
\label{rere}
\eta_{nn}>1-\sqrt{\frac{T_c}{T_h}}= \eta_{\rm CA}.
\EEA
Here $\eta_{\rm CA}$
is known as the Curzon-Ahlborn efficiency.  For
$u$ and $v$ satisfying to the work-extraction condition (\ref{mushi}), 
inequality (\ref{rere}) may or may not be satisfied. 

However, below in (\ref{galop}) we show that if the maximal work (over $u,v$) 
is extracted, (\ref{rere}) always holds.

Note that according to the Clausius formulation of the second law, if
there is a heat exchange between two thermal systems, the heat goes from
the hotter system to the colder one. Thus the temperature of the
initially hotter system is always larger or equal than the one of the
initially colder system.  Now recall that the Clausius formulation in
the present setup always holds globally in the sense that after the
engine cycle is completed, the hotter thermal bath (attached to R)
always has lost heat, while the colder bath (attached to S) always has
gained heat; see (\ref{oslo}, \ref{g_lobo}).  Condition (\ref{hub})
tells that the heat goes from hot to cold not only globally (i.e., for
the overall cyclic process of the engine functioning), but also locally,
in the work-extraction stage.  Indeed, the inverse condition $T_h'<T_c'$
will mean that the initially colder system got hotter at the end of the
work-extracting stage.  Thus according to (\ref{rere}) the maximal work
extraction is related to the local version of the Clausius formulation. 

Recall from (\ref{shushi}) that
for the work extraction at fixed spectra we need a finite difference
between the temperatures.  Condition (\ref{hub}), which is related to
the maximization of work, shows that the final temperatures $T_h'$ and
$T_c'$ have also to be different, in contrast to the classic
thermodynamic case, where after the maximal work-extraction the overall
system has one single temperature \cite{landau,callen}. 

\begin{figure}
\includegraphics[width=7cm]{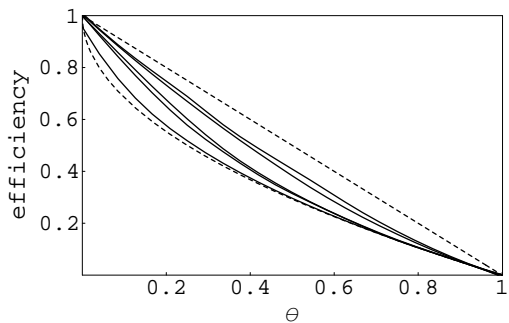}
\caption{
Normal curves: efficiency $\eta_{nn}$ at the maximal work  versus the
temperature ratio $\theta=\frac{T_c}{T_h}$ for various values of the
energy level number $n$. From bottom to top: $n=2,\, 501,\, 1001,\,
10001, 100001$.  Upper dashed line: Carnot efficiency $\eta_{\rm
Carnot}=1-\theta$.  Lower dashed line: Curzon-Ahlbron efficiency
$\eta_{\rm CA}=1-\sqrt{\theta}$.  It is seen that $\eta_{\rm Carnot}>
\eta_{nn} >\eta_{\rm CA}$. 
}
\hfill
\label{fig_1}
\end{figure}

\subsection{Maximization of work over the parameters $u$ and $v$ and
efficiency at maximal work.}
\label{maxw}

It remains to maximize the work $W_{nn}(u,v)$ over the parameters $u$ and $v$:
\BEA
W_{nn}= {\rm max}_{\,u,v}[\, W_{nn}(u,v)  \,].
\label{montr}
\EEA
Differentiating $W_{nn}(u,v)$ we
get the following equations for $\u$ and $\v$:
\BEA
\label{comrad1}
&&\frac{1+(n-1)\u}{1+(n-1)\v}
=\sqrt{\theta}\,\sqrt{\frac{\u}{\v}},\\
&&(1-\frac{\v}{\u})\,\frac{1+(n-1)\u}{1+(n-1)\v}
=\ln\frac{1}{\u}-\theta\,\ln\frac{1}{\v},
\label{comrad2}
\EEA
where
\BEA
\theta\equiv \frac{T_c}{T_h}\leq 1.
\EEA

\begin{figure}
\vspace{0.5cm}
\includegraphics[width=7cm]{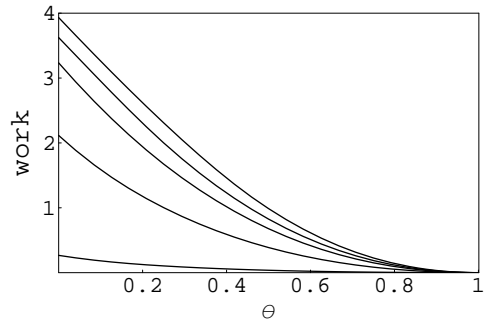}
\caption{
The dimensionless maximal work $\frac{W_{nn}}{T_h}$ versus the
temperature ratio $\theta=\frac{T_c}{T_h}$ for various values of the
energy level number $n$. From bottom to top: $n=2,\, 51,\, 251,\,
501, 1001$.   
}
\hfill
\label{fig_2}
\end{figure}

We see that $\u$ and $\v$|and thus $\frac{W_{nn}(\u, \v)}{T_h}$ and the
efficiency $\eta_{nn}(\u, \v)$ at the maximal work|depend only on the ratio
$\theta$ and the number $n$ of the energy levels. 

For $\theta=1$, Eqs.~(\ref{comrad1}, \ref{comrad2}) lead to $\u=\v$, 
and thus to $W_{nn}=\eta_{nn}=0$, as expected. For a fixed
spectrum the efficiency $\eta$ would not depend on $\theta$ at all. This changes
upon maximization of $W$:
The behavior of $\eta_{nn}$ as a function of
$\theta=T_c/T_h$ is shown in Fig.~\ref{fig_1}.  While
the Carnot efficiency $\eta_{\rm Carnot}=1-\theta$ is always an upper
bound for $\eta_{nn}$,  
it appears that $\eta$ also has a definite lower
bound given by the Curzon-Ahlborn efficiency:
\BEA
\label{galop}
\eta_{nn}\equiv \eta_{nn}(\u, \v)
>\eta_{\rm CA}
\equiv 1-\sqrt{\theta}.
\EEA
In particular, $\eta_{nn}$ converges toward $\eta_{\rm
CA}$ for $T_h\to T_c$; see next section and
Fig.~\ref{fig_1}. For small values of $n$,
$\eta_{nn}$ is closer to $\eta_{\rm CA}$ than to
$\eta_{\rm Carnot}$. For larger values of $n$ and for $\theta$ not very
close to $1$, $\eta_{nn}$ is closer to $\eta_{\rm
Carnot}$.  

In addition, Fig.~\ref{fig_1} shows that $\eta_{nn}$ (for a fixed $\theta$)
monotonically increases with the number $n$ of energy levels. A similar
monotonous increase holds for the dimensionless maximal work
$W_{nn}/T_h$; see Fig.~\ref{fig_2}.

\begin{figure}
\includegraphics[width=7cm]{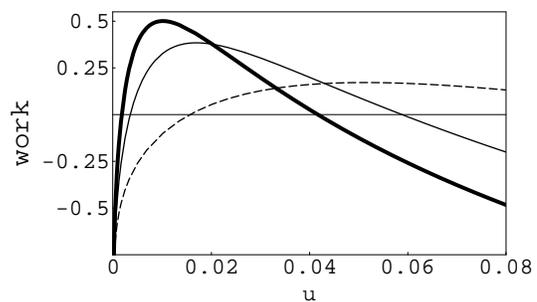}
\caption{The dimensionless work $\frac{W_{nn}(u, \v)}{T_h}$ versus the parameter $u$. The
parameter $v$ is set at its optimal value $\v$
that provides the unconditional maximum of $\frac{W_{nn}}{T_h}$. Here $T_c/T_h=0.5$.
Thick curve: $n=100$. Normal curve: $n=50$. Dashed curve: $n=10$. It is seen that the maximum of
$\frac{W_{nn}}{T_h}$ gets larger and sharper for larger values of $n$. The same qualitative conclusion is
kept for $\frac{W_{nn}(\u, v)}{T_h}$.
}
\hfill
\label{fig_11}
\end{figure}

Fig.~\ref{fig_11} presents a projection of the landscape of the
dimensionless work $\frac{W_{nn}(u,v)}{T_h}$.  We see that upon
increasing $n$, the maximum of $\frac{W_{nn}}{T_h}$ not only gets
larger, but it also becomes sharper.  (For increasing the visibility of
this effect, $\frac{W_{nn}}{T_h}$ is presented as a function of one
parameter $u$, with another being fixed at its optimal value; other
reasonable ways of taking projections around the maximum lead to similar
qualitative conclusions.) Imagine that various engines with different
characteristics $u$, $v$ and $n$ operate between the two thermal baths.
Now the engine with the best characteristics|i.e, larger $n$ and with
$u$ and $v$ closer to their optimal values for the given
temperatures|will produce the largest amount of work, and will thus
over-dominate the others.

\subsection{Special parameter windows}

\subsubsection{Equilibrium limit}

For the nearly equilibrium
situation $T_h\approx T_c$ one introduces the small parameter
$\varepsilon$ in the following way
\BEA
\label{bosfor}
\theta\equiv\frac{T_c}{T_h}=1-\epsilon, \qquad \frac{\u}{\v}=1+\epsilon \omega,
\EEA
Using (\ref{bosfor})
in (\ref{comrad1}, \ref{comrad2}) and keeping terms $\sim {\cal O}(\epsilon)$, we get
\BEA
\label{kap1}
&&\v=e^{-2 \omega}, \\
&&\frac{1}{\omega}=\frac{1-(n-1)e^{-2 \omega}}{1+(n-1)e^{-2 \omega}}.
\label{kap2}
\EEA
Eq. ~(\ref{kap1}) expresses $\v$ via $\omega$, while (\ref{kap2}) is a transcendental equation
for $\omega$. As graphical construction shows, (\ref{kap2}) has a unique positive solution for 
$n\geq 2$. Combining (\ref{kap1}, \ref{kap2}) with (\ref{kora1}, \ref{kora2}) we get for 
the maximal work $W_{nn}$ and for the efficiency $\eta_{nn}$ at the maximal work:
\BEA
W_{nn}= 
\frac{\epsilon^2 \omega^2 e^{-2 \omega}}{(1+ne^{-2 \omega})^2}+{\cal O}(\epsilon^2),
\\
\eta_{nn}=\frac{\epsilon}{2}+{\cal O}(\epsilon^2),
\label{joker}
\EEA
where $\omega>0$ is the solution of (\ref{kap2}). 
As (\ref{joker}) shows, in the considered order $\eta_{nn}$ 
coincides with the Curzon-Ahlborn efficiency: $\eta_{\rm
CA}=1-\sqrt{1-\epsilon}=\epsilon/2$. 

\begin{table}
\caption{
The maximal work $W_{23}$ and the relative difference 
$\frac{\eta_{23}-\eta_{\rm CA}}{\eta_{\rm CA}}$ between the efficiency 
at the maximal work and the Curzon-Ahlborn efficiency for various temperatures. Here
as in (\ref{23}),
the hot system R has two energy levels, while the cold system S has three energy levels.
}
\begin{tabular}{|c||c|c|}
\hline
\,\,  \,\,                     & $W_{23}$       & $\frac{\eta_{23}-\eta_{\rm CA}}{\eta_{\rm CA}}$  \\
\hline
\,\, $T_h=2$~~~ $T_c=1$   \,\, & \,0.0891062\, &\,0.0292807\,  \\
\hline
\,\, $T_h=3$~~~ $T_c=1$   \,\, & \,0.265928\, &\,0.0560355\,  \\
\hline
\,\, $T_h=10$~~~$T_c=1$   \,\, & \,1.97317\, &\,0.112379\,  \\
\hline
\end{tabular}
\label{tab1}
\end{table}

\subsubsection{Macroscopic limit: work and efficiency.}
\label{macro_macro}

If S and R are macroscopic, then $\ln n$ is large:
\BEA
\label{macro}
\ln n\gg 1.
\EEA
Indeed for a system containing $N\gg 1$ particles, the number of energy
levels scales as $e^{{\rm const}\,N}$.

Solution of Eqs.~(\ref{comrad1}, \ref{comrad2}) 
is found via expanding over the large parameter $\ln [n-1]$:
\BEA
\label{sa1}
&&\u=\frac{\theta (1-\theta)\ln [n-1]}{n-1}-
\frac{\phi+1-\theta^2}{n-1},\\
&&\v=\frac{\theta}{(1-\theta)\,[n-1]\,\ln [n-1]}+
\frac{\phi-1+\theta^2}{(1-\theta)^2(\,\ln[n-1]\,)^2},\nonumber\\
&&\label{sa2}\\
&&\phi\equiv \theta\ln\left[\theta (1-\theta) \ln [n-1]\right]
+\theta^2\ln\left[\frac{1-\theta}{\theta} \ln [n-1]\right].\nonumber\\
&&\label{sa3}
\EEA
The first terms in the RHS of (\ref{sa1}) and ( \ref{sa2}), respectively, are the
dominant ones, while the second terms are the corrections of the relative order
${\cal O}\left(\frac{1}{\ln[n-1]}\right)$.

\begin{figure}
\vspace{0.2cm}
\includegraphics[width=7cm]{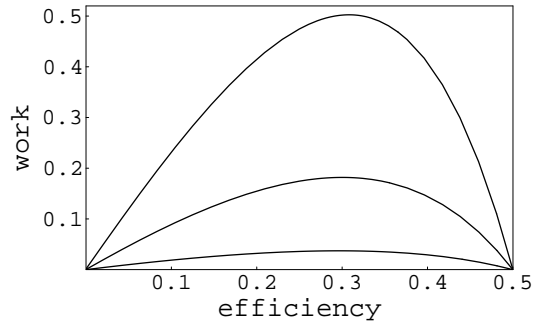}
\caption{
The dimensionless work $\frac{W_{nn}(\eta)}{T_h}$ defined in (\ref{monte})
versus the efficiency $\eta$.
Recall from (\ref{monte}) that $W_{nn}(\eta)$ is obtained by maximizing 
the work over all parameters
for a fixed efficiency $\eta$. 
For this figure $\theta=\frac{T_c}{T_h}=0.5$,
and from bottom to top: $n=2,\, 51,\, 101$. Note that
$\frac{W_{nn}(\eta)}{T_h}$ turns to zero two times: for $\eta=0$ 
and for $\eta=\eta_{\rm Carnot}=1-\theta$. The maximum of 
$\frac{W_{nn}(\eta)}{T_h}$ corresponds to the overall 
maximization of work; see Fig. ~\ref{fig_2}.
}
\hfill
\label{fig_3}
\end{figure}

Substituting (\ref{sa1}, \ref{sa2}) into (\ref{kora1}, \ref{kora2}) 
we get for the maximal work $W_{nn}$
and for the corresponding efficiency $\eta_{nn}$
\BEA
\label{be1}
\frac{W_{nn}}{T_h}&=&(1-\theta)\ln [n-1]-\frac{\phi+1+\theta^2}{\theta} \\
\label{be11}
&+&{\cal O}\left(\frac{1}{\ln[n-1]}\right),\\
\label{be2}
\eta_{nn}&=&1-\theta-\frac{2\theta}{\ln[\,n-1\,]}\ln [\,(1-\theta)\ln \left[\,n-1\,\right]\,]\\
&+&{\cal O}\left(\frac{1}{(\ln[n-1])^2}\right),
\label{be22}
\EEA
where $\phi$ is given by (\ref{sa3}), while the factors ${\cal
O}\left(\frac{1}{\ln[n-1]}\right)$ and ${\cal
O}(\frac{1}{(\ln[n-1])^2})$ in (\ref{be11}) and (\ref{be22}) can be
recovered with help of the correction in (\ref{sa1}) and ( \ref{sa2}),
respectively. 

It is seen from (\ref{be1}, \ref{be2}) that the efficiency converges to
the Carnot value, while the maximal work converges to the simple
expression 
\BEA
W_{nn}=T_h(1-\theta)\ln n, 
\EEA
which is a characteristic input thermal energy $T_h\ln n$ times the
Carnot efficiency $1-\theta$.  The corrections to these results|i.e.,
the last terms in the RHS of (\ref{be1}, \ref{be2})|are important as
well. They show that besides (\ref{macro}) one should
satisfy
\BEA
\label{macro1}
(1-\theta)\ln n\gg 1,
\EEA
and that due to negativity of the corrections 
the actual values of $W_{nn}$ and $\eta_{nn}$
increase in the macroscopic limit.

\subsubsection{Macroscopic limit: structure of the engine.}

An additional implications of the macroscopic limit is that the
temperatures $T_h'$ and $T_c'$ defined in (\ref{brams}) converge to
$T_h$ and $T_c$, respectively.  This is clear from (\ref{be2}) and from
$\frac{T_c'}{T_c}=\frac{T_h}{T_h'}=\frac{1-\eta_{nn}}{\theta}$, which
follows from (\ref{brams}).

As follows from (\ref{macro1}), the macroscopic limit is not compatible 
with the equilibrium limit $T_h\to T_c$.  An additional obstacle for
the macroscopic limit of the maximal work (but not of the efficiency) is
$\theta\to 0$, since then the correction $\frac{1}{\theta}$ in the LHS
of (\ref{be1}) diverges. 

For $\theta\to 1$ (or $T_c\to T_h$) both
the work and efficiency go to zero.  However, according to (\ref{macro1})
they tend to zero after having left the macroscopic regime
(\ref{be1}--\ref{be22}). For this regime (\ref{macro1}) implies that the
relative difference $1-\theta$ should be smaller than $\frac{1}{\ln n}$,
where ${\ln n}$ has the same order of magnitude as the number of
particles.

Note from (\ref{sa1}, \ref{sa2}) and from (\ref{bzez4}) that the energy
gaps $\eps_2$ and $\mu_2$ scale as $\ln n$, as should be for a
macroscopic systems.  Another aspect is uncovered when
comparing (\ref{sa1}, \ref{sa2}) with (\ref{bzez3}, \ref{bzez4}): in the
limit (\ref{macro}) for the high-temperature system $\R$ the population
is concentrated on the higher ($n-1$ times degenerate) energy level
$\eps_2$, while for the low-temperature system $\S$ the population is
concentrated on the lowest ($\mu_1=0$) energy level. In particular, this
means that for both $\S$ and $\R$ in the limit (\ref{macro}), the
canonical ensemble is equivalent (in the leading order of $\frac{1}{\ln
n}$) to the micro-canonic ensemble, where all the population is put on
one [possibly (quasi)degenerate] energy level. This is what one
expects for a macroscopic system.  For the spectrum 
(\ref{go1}, \ref{go2}), the equivalence of the canonic and micro-canonic
ensembles holds under special conditions only, and those conditions are
realized after the maximization of the work, as shown by (\ref{sa1},
\ref{sa2}).

\subsubsection{System with $n\not=m$.}

So far we studied the case when both subsystems S and R have the same
size, i.e. the same number $n=m$ of energy levels.  
One may want to know what
happens for $n\not=m$. One of the simplest models of this type is
\BEA
\label{23}
n=2\qquad {\rm and} \qquad m=3. 
\EEA

While for $m=n$ we had the conditions (\ref{go1}, \ref{go2}), which
considerably simplified the subsequent analysis, no analog of these
conditions holds for (\ref{23}). Thus we have to rely on numerical
investigations.  It appears that we can keep all qualitative conclusions;
see Table I. 

{\it i)} The dimensionless maximal work $\frac{W_{nm}}{T_h}$ is still a
function of the temperature ratio $\theta=T_c/T_h<1$. {\it ii)} The
optimal values of the energy spacings|though not satisfying any simple
condition similar to (\ref{go1}, \ref{go2})|still depend only on
$\theta$. {\it iii)} The efficiency at the maximal work $\eta_{nm}$ is still
bounded from below by the Curzon-Ahlborn value $1-\sqrt{\theta}$ and
from above by the Carnot value $1-\theta$.

\subsection{Conditional Maximum}

When seeking the maximum of $W$ over the spectra one may impose
additional constraints. We study two such scenarios below.

\subsubsection{Maximization of work for fixed efficiency.}

Consider the maximum of $W_{nn}(u,v)$ in
(\ref{kora1}), when the maximization over the parameters $u$ and $v$ is
carried out under the condition of a given efficiency (\ref{kora2}). This 
defines the dimensionless maximal work $\frac{W_{nn}(\eta)}{T_h}$
for a fixed efficiency $\eta$:
\begin{gather}
\frac{W_{nn}(\eta)}{T_h}
={\rm max}_{u}\left[\, 
\frac{   (n-1)\,\eta\,[u-u^{ \frac{1-\eta}{\theta} }  ]  \ln\frac{1}{u}
}{(1+(n-1)u)(1+(n-1)u^{\frac{1-\eta}{\theta}})}
\, \right].
\label{monte}
\end{gather}
Note that $W_{nn}(\eta)$ is still a function of the temperature ratio $\theta=\frac{T_c}{T_h}$.

As seen in Fig.~\ref{fig_3}, $\frac{W_{nn}(\eta)}{T_h}$ as a function of
$\eta$ is a bell-shaped curve, which turns to zero once for $\eta=0$ and
then for $\eta=\eta_{\rm Carnot}$; recall the discussion in section
\ref{carnot_finite}. Fig.~\ref{fig_3} shows that
$\frac{1}{T_h}W_{nn}(\eta)$ increases upon increasing the number of
energy levels, while the maximum $\eta_{nn}$ of
$\frac{1}{T_h}W_{nn}(\eta)$ shifts toward larger values, 
since in the macroscopic limit $\frac{1}{T_h}W_{nn}(\eta)$
will have its maximum very close to the Carnot value $\eta=1-\theta$;
recall (\ref{be2}). 

\subsubsection{Emergence of the Curzon-Ahlborn limit during partial
optimization of work: An example}

It may be instructive to follow in more detail, but via a particular
example, how the Curzon-Ahlborn lower bound for the efficiency emerges
during a constrained maximization. Assume that the systems R and S have,
respectively, the spectra $\{ a_{\rm R}\,\eps_k\}_{k=1}^n$ and $\{ a_{\rm
S}\,\eps_k\}_{k=1}^n$, which differ only by positive scaling factors $a_{\rm
R}$ and $a_{\rm S}$.  The initial states of R and S are given by
(\ref{1}).  We shall not optimize over the interaction with the
work-sources, but rather impose the SWAP transformation (\ref{duko}) for
the work-extraction. The work extracted during this transformation is
deduced from (\ref{work}):
\BEA
\label{kram1}
W_{nn}(a_{\rm R}, a_{\rm S})=(a_{\rm R}-a_{\rm S})\sum_{k=2}^n \eps_k (r_k-s_k),
\EEA
where $r_k=\frac{ e^{-\beta_h a_{\rm R} \eps_k  } }{  \sum_{l=1}^n e^{-\beta_h a_{\rm R} \eps_l  }}$ and
$s_k=\frac{ e^{-\beta_h a_{\rm S} \eps_k  } }{  \sum_{l=1}^n e^{-\beta_h a_{\rm S} \eps_l  }}$.
In the second step the systems R and S are subjected to free relaxation, as described
in section \ref{efo}. Without loss of generality we assume $\frac{T_c}{T_h}<
\frac{a_{\rm S}}{a_{\rm R}}$, which means that R (S) has higher (lower) temperature
\footnote{Since we did not optimize over the interaction with the sources of work,
$W_{nn}(a_{\rm R}, a_{\rm S})$ in (\ref{kram1}) can be negative, e.g., for $T_h=T_c$,
which means that the work is put into the system.}.
The efficiency of work-extraction is obtained from (\ref{kram1}, \ref{am3}, \ref{ef}):
\BEA
\label{grimm}
\eta_{nn}(a_{\rm R}, a_{\rm S})=1-\frac{a_{\rm S}}{a_{\rm R}}.
\EEA

Note that the duration of the work extracting cycle is finite. 
Instead of doing the full optimization over $a_{\rm S}$ and $a_{\rm R}$,
we now employ a necessary conditions 
$\frac{\partial W_{nn}}{\partial a_{\rm
R}}=\frac{\partial W_{nn}}{\partial a_{\rm S}} =0$,
for the work (\ref{kram1}) to be optimal over these two parameters. 
Working out $\frac{\partial W_{nn}}{\partial a_{\rm
R}}+\frac{\partial W_{nn}}{\partial a_{\rm S}} =0$
we get
\BEA
\label{gera}
1-
\frac{a_{\rm S}}{a_{\rm R}}=1-\sqrt{\frac{T_c C_c}{T_h C_h}},
\EEA
where $C_h=\beta^2_h a^2_{\rm R} [\langle \eps^2\rangle_h-\langle
\eps\rangle^2_h] $ and $C_c=\beta^2_c a^2_{\rm S} [\langle
\eps^2\rangle_c-\langle \eps\rangle^2_c] $ are the equilibrium specific heats
of R and S, respectively, calculated for the initial states
at the temperatures $T_h$ and $T_c$ (recall that
$T_h>T_c$). 

For a finite number of energy levels $n$, the equilibrium specific heat
$C$ is a non-monotonous function of temperature: for low temperatures
$C$ naturally goes to zero, while for for very large temperatures it is
zero again, since there is a maximal energy a finite-level system can
have.  Thus for a finite $n$, the ratio $\frac{C_c}{C_h}$ may be larger
or smaller than one, so that (\ref{gera}) does not imply any definite
relation between the actual efficiency (\ref{grimm}) and the
Curzon-Ahlborn value. However, for $n=\infty$ the equilibrium specific
heat can be [but need not be] a monotonically increasing function of the
temperature, since now the system can accept infinitely large energies.
An example of such a behavior is given by a harmonic oscillator with
frequency $\omega$, where $\eps_n=\hbar\omega n$, and the specific heat
is $C=b^2 e^{b}(e^b-1)^{-2}$, with $b=\hbar\omega /T$.  Thus, now
$\frac{C_c}{C_h}<1$, since $T_h>T_c$, and (\ref{gera}) implies that the
Curzon-Ahlborn value is a lower bound for the efficiency (\ref{grimm}). 

We draw two conclusions: {\it i)} upon
partial maximization the Curzon-Ahlborn value may or may not be a lower
bound for the efficiency depending on the system details; {\it ii)} the
emergence of the Curzon-Ahlborn lower bound is facilitated for systems
with classical (i.e., unlimited from above) spectrum, since now the equilibrium
specific heat can be an increasing function of the temperature.

\section{Optimizing the second step}
\label{opto_2}\label{V}\label{controlled_relaxation}

Now we want to extract work also in the second stage, i.e., during the
relaxation of R and S. In contrast to the previous consideration, here
we {\it impose} that the couplings of R and S
with their respective baths are weak.  We start at the initial state
(\ref{omega_in}) of S+R.  The maximal work-extracting transformation is
applied, the work (\ref{am2}) is extracted, and R (S) is left in the
final state $\rho_{\rm fin}$ ($\sigma_{\rm fin}$), as given by
(\ref{marginal_final}). Recall from (\ref{kasablanka}) that $\rho_{\rm
fin}$ and $\sigma_{\rm fin}$ commute with their Hamiltonians.  Due to
the weak-coupling assumption the overall initial state of R (S) and the
hot (cold) bath is, to a good approximation, factorized: $\rho\otimes
\rho_{\rm B}$ ($\sigma\otimes \sigma_{\rm B}$), where $\rho_{\rm B}$
($\sigma_{\rm B}$) is the equilibrium state of the hot (cold) bath.  Since the
interaction with the bath is not essential during the first [thermally
isolated] stage, the same factorization holds after this stage:
$\rho_{\rm fin}\otimes \rho_{\rm B}$ and $\sigma_{\rm
fin}\otimes \sigma_{\rm B}$.

If we just let R (S) to interact with the thermal bath at temperature
$T_h$ ($T_c$), it will relax back to the initial state $\rho$
($\sigma$). We can however control this relaxation process and extract
an additional amount of work.

According to the standard thermodynamic argument
\cite{landau,callen}, the {\it maximal} work extractable|via a
cyclic-Hamiltonian process|from a system in the initial state $\rho_{\rm
fin}$ in contact with the thermal bath at temperature $T_h$ is bounded
from above by the free energy difference
\BEA
\Delta F_\R& =& {\rm tr}[\rho_{\rm fin} H_\R]+T_h{\rm tr}[\rho_{\rm fin}\ln\rho_{\rm fin}]\nonumber\\
&-&{\rm tr}[\rho H_\R]-T_h{\rm tr}[\rho\ln\rho]\equiv\Delta U_\R-T_h\Delta S_\R,~~~\\
\Delta U_\R &\equiv & {\rm tr}[\rho_{\rm fin} H_\R] -{\rm tr}[\rho H_\R],\\
\Delta S_\R &\equiv & -{\rm tr}[\rho_{\rm fin}\ln\rho_{\rm fin}]+{\rm tr}[\rho\ln\rho],
\label{free}
\EEA
where $\rho$ is the equilibrium state (\ref{1}), and where $\Delta U_\R$
and $\Delta S_\R$ are the changes of the energy and the von Neumann
entropy, respectively. 

A detailed discussion of (\ref{free}) and its derivation is given in
Appendix \ref{app1}. Moreover, in Appendix \ref{app2} we show that if
the initial state commutes with the Hamiltonian, $[\rho_{\rm
fin},H_\R]=0$, as is the case according to (\ref{kasablanka}), the bound
(\ref{free}) is achieved via:\\
(1.) {\it Suddenly} changing the energy spacings of $H_\R$ according to (\ref{kagan2},
\ref{kagan3}, \ref{kagan33}). This brings R to the local equilibrium
with the hot bath, since at the end of the sudden change
the temperature of R is equal to the bath temperature $T_c$.
The change is sudden as compared to the relaxation time
induced by the weak coupling to the hot bath, so that the
interaction with the hot bath during the change is neglected. \\
(2.) {\it Slowly} changing the spacings back to their initial values. Now the change
is much slower than the relaxation time of R driven by the
weak-coupling with the hot bath. Thus R is always in the local
equilibrium with the hot bath; see the discussion around (\ref{tartar}).
The equilibrium state $\rho$ is attained at the end of this isothermal
process. The work-extraction in this stage takes a time much longer than the
relaxation time of R which [due to the weak-coupling assumption] is already
larger than internal times of R. 
Thus the overall process has a small work extraction power.

The heat received from the hot thermal bath is (compare with (\ref{am3})):
\BEA
Q_h=\tr (H_\R\,\rho)-
\tr (H_\R\rho_{\rm fin})+\Delta F_\R=-T_h\Delta S_\R,
\label{gazan1}
\EEA
where $\Delta F_\R$ is given by (\ref{free}).

The same procedure is applied to $\S$ thereby extracting the work $\Delta F_\S$:
\BEA
\Delta F_\S& =& {\rm tr}[\sigma_{\rm fin} H_\S]+T_c{\rm tr}[\sigma_{\rm fin}\ln\sigma_{\rm fin}]\nonumber\\
&-&{\rm tr}[\sigma H_\S]-T_c{\rm tr}[\sigma\ln\sigma]\equiv\Delta U_\S-T_c\Delta S_\S,~~
\label{free_S}
\EEA
where $\Delta U_\S$ and $\Delta S_\S$ are defined analogously to (\ref{free}):
\BEA
\Delta U_\S &\equiv & {\rm tr}[\sigma_{\rm fin} H_\S] -{\rm tr}[\sigma H_\S],\\
\Delta S_\S &\equiv & -{\rm tr}[\sigma_{\rm fin}\ln\sigma_{\rm fin}]+{\rm tr}[\sigma\ln\sigma],
\label{free_1}
\EEA

\begin{table}
\caption{The maximal value of $\W_{33}$, given by (\ref{oront}) with maximization
over the parameters $u$ and $v$, compared to
the maximal value of $W_{33}$; see (\ref{montr}). The maximization is done for $n=m=3$
(two three-level systems).
We also present the relative difference $\frac{\eta_{33}-\eta_{\rm CA}}{\eta_{\rm CA}}$
between
the efficiency $\eta_{33}$ at the maximal work $W_{33}$ and the Curzon-Ahlborn efficiency
$\eta_{\rm CA}=1-\sqrt{T_c/T_h}$. Recall that the work $\W_{33}$ is extracted with
the Carnot efficiency; see (\ref{gazan3}, \ref{gazan33}).
}
\begin{tabular}{|c||c|c|c|}
\hline
\,\,  \,\,                     & $W_{33}$       & $\frac{\eta_{23}-\eta_{\rm CA}}{\eta_{\rm CA}}$ & $\W_{33}$  \\
\hline
\,\, $T_h=2$~~~ $T_c=1.7$   \,\, & \,0.009271\, &\,0.000931\, & 0.0198908  \\
\hline
\,\, $T_h=2$~~~ $T_c=1$   \,\, & \,0.128486\, &\,0.0142578\,  &  0.478908  \\
\hline
\,\, $T_h=10$~~~$T_c=1$   \,\, & \,3.07425\, &\,0.0711658\,  & 9.20744  \\
\hline
\end{tabular}
\label{tab2}
\end{table}

The total work extracted during the cycle reads
\BEA
\label{gazan22}
\W_{nm}(\eps,\mu)&=&W_{nm}(\eps,\mu)+\Delta F_R+\Delta F_S\\
&=&-T_h\Delta S_\R-T_c\Delta S_\S,
\label{gazan2}
\EEA
where $W_{nm}(\eps,\mu)$ is the maximal work (\ref{marginal_work})
[optimized over the interaction with the sources of work] extracted
during the first stage.  Note that the only feature of
$W_{nm}(\eps,\mu)$ employed in passing from (\ref{gazan22}) to
(\ref{gazan2}) is that the evolution of R+S in the first stage is
themally isolated, i.e., $W_{nm}(\eps,\mu)$ is equal to the energy
difference of R+S; see (\ref{work}). 

The efficiency that follows from (\ref{gazan1}, \ref{gazan2}) is
\BEA
\label{gazan3}
\Ce_{nm}(\eps,\mu)=1+\frac{T_c}{T_h}\, \frac{\Delta S_\S}{\Delta S_\R}.
\EEA

The work (\ref{gazan2}) is now optimized over the spectra
$\ep_n,...,\ep_2$ and
$\mu_n,...,\mu_2$ of R and S, respectively, for fixed temperatures.

For $\R$ and $\S$ having the same number of energy levels: $n=m$, the
numerical optimization of (\ref{gazan2}) produced the same result
(\ref{go1}--\ref{go4}) as for the work extraction at a finite power: the
optimal work is achieved for the upper levels being $n-1$ times
degenerate. 

As we already know, Eqs.~(\ref{go1}, \ref{go2}) imply the factorized
final state (\ref{omega_fin}): $\rho_{\rm fin}=\sigma$ and $\sigma_{\rm
fin}=\rho$; see (\ref{duko}). Recalling the definitions (\ref{free},
\ref{free_1}) of $\Delta S_\S$ and $\Delta S_\R$, we get
\BEA
{\Delta S_\S}+{\Delta S_\R}=0.
\label{roma}
\EEA

The physical meaning of this condition becomes clear when noting that if
the factorization (\ref{duko}) was invalid|i.e. if there were
correlations in the final state (\ref{omega_fin})|the sum ${\Delta
S_\S}+{\Delta S_\R}$ would  be larger than zero, as implied by the
sub-additivity of the von Neumann entropy\footnote{If $S_{12}=-{\rm
tr}\rho_{1+2}\ln \rho_{1+2}$ is the von Neumann entropy of a composite
system with density matrix $\rho_{1+2}$, then the sub-additivity implies
\cite{delgado}: $S_{1+2}\leq S_1+S_2$, where $S_i=-{\rm tr}\rho_{i}\ln
\rho_{i}$, $i=1,2$, and where $\rho_1={\rm tr}_2 \rho_{1+2}$,
$\rho_2={\rm tr}_1 \rho_{1+2}$.}.
Since after the second step of the engine operation, the systems $\R$
and $\S$ return to their initial states (\ref{1}), the surplus entropy
${\Delta S_\S}+{\Delta S_\R}$ has to be consumed by the thermal baths,
thereby increasing their entropy. Thus, the factorization condition
(\ref{duko}) eliminates this potential channel for entropy generation. 

Eqs. (\ref{gazan3}) and (\ref{roma}) imply that the efficiency is equal
to the {\it maximally possible} Carnot value:
\BEA
\label{gazan33}
\Ce_{nn}=\eta_{\rm Carnot}=1-\frac{T_c}{T_h}.
\EEA

Using (\ref{bzez1}--\ref{bzez4}) we get for the work (\ref{gazan2})
\BEA
&&\W_{nn}(u,v)= (T_h-T_c)\left(\,{\rm tr}[\sigma\ln\sigma]-{\rm tr}[\rho\ln\rho]\,\right)~~\\
&&=(T_h-T_c) \left[\,
\ln\frac{1+(n-1)u}{1+(n-1)v}\right.\nonumber\\
&&\left.
+(n-1)\frac{v\ln v-u\ln u+(n-1)uv\ln\frac{v}{u}}{(1+(n-1)u)(1+(n-1)v)}\,
\right].~~~~
\label{oront}
\EEA

The latter expression is to be optimized over $u$ and $v$. 
It should be clear from our constructions that
\BEA
\W_{nn}\equiv {\rm max}_{u,v}[\,\W_{nn}(u,v)\,]
\geq W_{nn}.
\EEA
It is seen that $\frac{ \W_{nn} }{T_h}$|analogously to $W_{nn}$ in (\ref{montr})|is
a function of the temperature ratio $\theta\equiv\frac{T_c}{T_h}$.

The difference $\W_{nn}-W_{nn}$ quantifies how much work has been traded
in by sacrifying the power. As Table II shows, especially for
$T_h\approx T_c$, the ratio $\frac{\W_{nn}}{W_{nn}}$ can approach $\sim
20$. 

\subsection{Can one extract a finite amount of work at a finite power
and with a finite-level engine?}
\label{comrad_E}

Recall that in section \ref{carnot_finite} we discussed that when a
finite-level engine approaches the Carnot efficiency at a finite power,
the work extracted per cycle vanishes, though the cycle duration is
finite.  This is not anymore the case for a macroscopic engine, as we
saw in section \ref{macro_macro}.  Above we have shown that if one
imposes the zero-power condition, the work-extraction at the Carnot
efficiency is possible. 

Can one extract a finite amount of work at a finite power and with a
finite-level engine? In other words, is it possible to extract in the
second stage the amounts of work (\ref{free}, \ref{free_S}) without
employing very long processes? The answer is yes \cite{aa}. A finite
cycle time is achieved, since the coupling to the bath is kept weak
before and after the work-extraction process, but it is made finite in
the intermediate stages.  However, the scheme proposed in \cite{aa} has
a serious drawback: one should employ the sources of work-extraction
that act not only on $\R$ and $\S$, but also on the corresponding
thermal baths. The latters are normally out of any direct control, and
it is not clear to what extent the scheme proposed in \cite{aa} is
realistic.  Thus the question is open.

\section{Discussion and conclusions}
\label{VI}

\subsection{Summary.}

Our purpose was in deducing the physics of quantum heat engines from
maximizing the work [extracted per cycle] under various constraints.  We
have found that the work maximization can be introduced on three
different levels:

{\bf I.} Our model is described in section \ref{model}; see also Fig.~\ref{fig_0}.
One optimizes the extracted work over the interaction of the
quantum systems $\S$ and $\R$ with the external source of work. In
general, this procedure is not sufficient for extracting a finite amount
of work at $T_h\not=T_c$. In addition, the intermediate stages of the
engine functioning are not described by well defined temperatures: if $\S$
and $\R$ had well defined temperatures before the work-extraction, they are
not guaranteed to have definite temperatures after (or during) this
process. 

{\bf II.} One maximizes the work, in addition, over the spectral
structure of $\S$ and $\R$.  This turns out to be a crucial step,
since after this optimization one finds that {\it i)} $\S$ and $\R$ are
described by well defined temperatures in the intermediate stages of the
engine operation. {\it ii)} The Clausius formulation of the second
law|heat goes from higher to lower temperature|is satisfied not only for
the total work-extracting cycle [as it should], but also locally for the
intermediate stages of the process.  Thus the local thermodynamic
structure emerges as a result of work optimization.  {\it iii)} The
efficiency at the maximal work is bounded from {\it below} by the
Curzon-Ahlborn efficiency $1-\sqrt{T_c/T_h}$. This is in addition to the
upper Carnot bound $1-T_c/T_h$, which holds for this model generally,
i.e., with or without any optimization. The important feature of these
limits is that they are completely system-independent. \\ 
Note that the duration of the engine cycle is finite.
Thus the power of work is finite and [within our model]
tunable to a large extent; see section \ref{power} for details. 

{\bf III.} The Curzon-Ahlborn bound is approached close to equilibrium
$T_h\to T_c$, while the Carnot upper bound is reached|at a finite
power|for a large system $\S$ and $\R$. In the latter macroscopic limit
$\S$ and $\R$ (tuned to extract the maximal work) have the expected
features of macroscopic systems, e.g., canonic and micro-canonic
ensembles are equivalent for them. The equilibrium and macroscopic
limits do not commute. 

{\bf IV.} Finally, one optimizes the extracted work also over the
interaction of $\S$ and $\R$ with their respective thermal baths.  This
optimization was carried out via increasing the cycle duration and thus
reducing the power of work extraction. The main result is that the
Carnot value for the efficiency|with a finite amount of work extracted
at a small power|is reached for finite systems $\S$ and $\R$. Thus there
are two options for operating close to the Carnot efficiency: either one
extracts a finite amount of work per a very long cycle, or a small
amount of work per a finite cycle.

\subsection{Relation to previous work}

\subsubsection{Curzon-Ahlborn efficiency.}

The Curzon-Ahlborn efficiency has repeatedly been found as the
(approximate) efficiency of model macroscopic heat engines functioning
at a finite power \cite{CA,leff}.  Ref.~\cite{kosloff} studies a quantum
heat engine model, and derives the Curzon-Ahlborn value, but in the limit of
vanishing interaction with sources of work.  Taking up the early
approach of \cite{CA}, it was stated recently that the Curzon-Ahlborn
value is an {\it upper} bound for the efficiency of a heat engine
operating at the maximum power and according to the rules of linear
transport theory \cite{CA_van}. 

In contrast, we show that the Curzon-Ahlborn value is the {\it lower}
bound for the efficiency at the maximal work and a finite power of work.
This clarifies the meaning of the Curzon-Ahlborn value, since our
derivation does not employ the linear transport theory, and is based on
the fundamental level of quantum mechanics. It is important that rather
different physical mechanisms can lead to the same expression of the
Curzon-Ahlborn value. 

\subsubsection{Carnot efficiency.}

According to standard thermodynamics the efficiency of any heat engine
is bounded by the Carnot value \cite{landau,callen}. Although this
result was derived within (nearly)equilibrium thermodynamics, it can be
extended to a rather general class of non-equilibrium heat engines; see
(\ref{oslo}, \ref{carnot}) and \cite{jmod}. While useful as an upper
bound, the Carnot efficiency by itself is often considered to lack
practical significance, since the the power of work goes to zero in this
limit.  This has been a motivation to develop the so-called finite-time
[finite-power] thermodynamics \cite{finite_time}. 

The quantum situation can be even more restrictive in this context,
since typical implementations are based on the Otto cycle, for which not
only the power, but also the extracted work goes to zero when
approaching the Carnot bound \cite{nori,ki,2}. 

For a finite-level quantum heat engine operating close to the Carnot
efficiency we have two related results. First we show that for such an
engine operating with a finite cycle duration, the work extracted per
cycle goes to zero when the efficiency approaches the Carnot value; see
our discussion at the end of section \ref{la_Habana}. Second result is
that the same quantum engine can extract a non-zero amount of work|but
with a long cycle duration| sharply at the Carnot efficiency, provided
one properly optimizes also over the second step, the bath-engine
interaction. 

Results on small quantum \cite{60,linke,linke_talk,nori_c,beretta} or
classical \cite{saka,astumian} engines working at zero power, but with
the Carnot efficiency were already reported in literature. Most of these
results \cite{60,linke,linke_talk,saka,astumian} concern continuously
[not cyclically] working engines, where the external source of work is
absent and the work is performed by moving particles against an external
force.  In contrast, Refs.~\cite{nori_c,beretta} report on a cyclically
operating quantum heat engine that [at a small power] deliver with the
Carnot efficiency a finite amount of work per cycle to an external
source of work. Both these works attempt to model quantum Carnot cycle. 

We also show that the macroscopic engine can extract a finite amount of
work at a finite-power and with [almost] the Carnot efficiency.  To our
knowledge this effect was so far never reported in the literature. 

\subsection{Open questions.}

We leave open two important questions: 

(1.) How the engine tunes itself to the maximal work-extraction regime?
Some hints on this point were expressed at the end of section
\ref{maxw}. 

(2.) Can one extract work at a finite power by employing a finite-level
engine operating at the Carnot efficiency? This question was discussed
in section \ref{comrad_E}. 

\acknowledgements The work was supported by Volkswagenstiftung grant
``Quantum Thermodynamics: Energy and information flow at nanoscale''.
RSJ acknowledges support jointly from Max Planck Institute for
Solid State Physics, Stuttgart and Institute for Theoretical Physics I,
University of Stuttgart.  We thank M. Henrich, Th. Jahnke, and G.
Reuther for valuable discussions.

\appendix

\section{Collisional relaxation.}
\label{app_collisions}

The purpose of this discussion is to outline a realistic example of a
finite-duration [collisional] relaxation process which is consistent
with the thermally isolated work extracting setup described in section
\ref{first_step}. Our presentation follows to \cite{partovi,vaks,mito}. 

The thermal bath is modeled as a collection of $N\gg 1$ independent
equilibrium systems (particles) with density matrices
$\omega_{i}=\frac{1}{Z_i}\exp[-\beta H_{i}]$ and Hamiltonians $H_{i}$,
where $i=1,..,N$, and where $1/\beta=T$ is the bath temperature. This
formalizes the intuitive notion of the bath as a collection of
many weakly-interacting equilibrium systems. 

The target system R starts in [an arbitrary] initial state $\rho$ and has
Hamiltonian $H_{\rm R}$. The collisional relaxation is realized when the
particles of the bath sequentially interact [collide] with R. Consider the first 
collision. The initial state of R and the first bath particle is
$\Omega_{1+\R}=\rho\otimes\omega_1$, the interaction between them is realized via a unitary
operator ${\cal V}$, so that the final state after the first collision is 
$\Omega_{1+\R}'={\cal V}\Omega_{1+\R}{\cal V}^\dagger$. Define separate final states:
\BEA
\rho'={\rm tr}_1 \Omega_{1+\R}', \qquad
\omega_1'={\rm tr}_\R \Omega_{1+\R}',
\EEA
where ${\rm tr}_1$ and ${\rm tr}_\R$ are the partial over the first particle and R, respectively.
Recall the definition (\ref{relative_entropy}) of the relative
entropy. The unitarity of ${\cal V}$ implies
\BEA
S[\,\Omega_{1+\R}'\, ||\, \rho'\otimes \omega_1\, ]&=&
{\rm tr}[\Omega_{1+\R}\ln \Omega_{1+\R}
]\nonumber\\
&-&{\rm tr}[\Omega_{1+\R}'\ln(\rho'\otimes \omega_1)].
\label{ortega}
\EEA
Employing $\omega_{1}=\frac{1}{Z_1}\exp[-\beta H_{1}]$ and
$S[\,\Omega_{1+\R}'\, ||\, \rho'\otimes \omega_1\,]\geq 0$ 
in (\ref{ortega}) we get
\BEA
\label{kabanets}
T \Delta S_{\R}+\Delta U_1\geq 0,
\EEA
where $\Delta S_{\R}={\rm tr}\left [-\rho'\ln \rho' + \rho\ln \rho 
\right]
$ and $\Delta U_{1}={\rm tr}(H_1\left  [\omega_1'-\omega_1\right])$
are, respectively, the change of the entropy of R and the average energy of 
the first particle. 

We now require that the that interaction ${\cal V}$ conserves the
energy: $\Delta U_1=-\Delta U_R$. Using this in (\ref{kabanets}) one has
\BEA
\label{kabanets1}
\Delta U_\R -T \Delta S_{\R}\leq 0. 
\EEA
Since we did not use any special feature of the initial state of R,
(\ref{kabanets1}) holds for subsequent collisions of R with the bath
particles. Thus $\Delta U_\R -T \Delta S_{\R}$ decays in time, and it should
attain its minimum. It is well-known \cite{landau} that this minimum is
reached for the Gibbs matrix $\rho\propto e^{-\beta H_\R}$: the
collisions drive $\R$ to equilibrium starting from an arbitrary state.
The coupling with the bath is switched on during the collision only,
but since this coupling can be sizable, the relaxation time can be 
very short \cite{partovi}.

Further results and concrete scenarios of collisional relaxation are
given in \cite{partovi,vaks,mito,ziman}. This includes the rate of
[exponential] convergence to equilibrium which was favourably
compared to experiments in \cite{vaks}. 

As for the implementation of the work extracting pulse described
in section \ref{first_step}, one notes that within the present
relaxation model, the duration of the pulse has to be shorther than the
time between the collisions. If the latter time is much larger than both
the collision time and the pulse time, almost any implementation of the
pulse will amount to a thermally isolated process. 

\section{System with $n-1$ fold degenerate spectrum.}
\label{lion}

Consider an $n$-level quantum system R whose upper $n-1$ energy levels
are degenerate; see (\ref{go1}).  Let us show that the behavior of
$\R$ is equivalent to that of a two-level system ($n=2$) with respect to
all transformations that do not resolve the $n-1$ fold degeneracy of the
spectrum (\ref{go1}). To this end we introduce for $n=3$
generalized Pauli matrices [generalizing to $n>3$ is straightforward]
\BEA
\hat{\sigma}_3=
\left(\begin{array}{rrr}
1 & 0 & 0 \\
0 & 1 & 0 \\
0 & 0 & -1
\end{array}\right),\,
\hat{\sigma}_1=
\left(\begin{array}{rrr}
0 & 0 & 1 \\
0 & 0 & 1 \\
1 & 1 & 0
\end{array}\right),\,
\hat{\sigma}_2=i\hat{\sigma}_1\hat{\sigma}_3.~~~
\EEA
It is clear that {\it i)} the algebra of these matrices is identical to
that of the corresponding Pauli matrices; {\it ii)} Any perturbation of
the Hamiltonian $H_{\rm R}$ of R by an arbitrary linear combination of
$\{\hat{\sigma}_i\}_{i=1}^3$ does not change the double-degeneracy of
$H_{\rm R}$; {\it iii)} If the initial state of R can be expressed only via
$\{\hat{\sigma}_i\}_{i=1}^3$, then during any interaction with another quantum
system|which is written only via $\hat{\sigma}_i$ and arbitrary
operators of this system|R is dynamically equivalent (via the
Heisenberg representation) to a two-level system. An example
of such an initial state is when $\R$ is described by a
definite temperature.

\section{}
\subsection{Free energy bound for the maximal work extractable
from a system in contact with a bath. }
\label{app1}

We are given a quantum system in a non-equilibrium initial state $\Phi$ and
Hamiltonian $H$.  Some work is to be extracted from this system via a
cyclic-Hamiltonian, 
\BEA
H(0)=H(\tau)=H, 
\EEA
thermally isolated process. We want
to give an upper bound for this work.  The
work-extraction process is unitary, and thus it conserves the
eigenvalues of $\Phi$.  In particular, it conserves the von Neumann
entropy
\BEA
S[\Phi]=-{\rm tr}\, [\Phi\ln \Phi].
\EEA

Recall that the (positive) extracted work is defined as the difference
between the initial and final energies. We can give an upper
bound for the maximal extractable work by looking for a final state with
the {\it minimal} energy compatible with the above conservation of the
von Neumann entropy \cite{landau,callen}.  The corresponding minimization
procedure is standard in statistical physics \cite{landau}|since it is
dual to the maximization of the entropy for a fixed average energy|and
it produces for the final state a Gibbsian density matrix
\BEA
\Phi(\beta_{\rm f})=\frac{1}{Z}\, e^{-\beta_{\rm f} H}, \qquad Z={\rm tr}\, e^{-\beta_{\rm f} H},
\label{kalimantan}
\EEA
with an inverse final temperature $\beta_{\rm f}$
defined via the entropy conservation:
\BEA
\label{ga2}
S[\Phi]=S[\Phi(\beta_{\rm f})].
\EEA
The upper bound for the maximal work is now
\BEA
\label{ga4}
W_{\rm th}={\rm tr}\, [H\Phi]
-{\rm tr}\, [H\Phi(\beta_{\rm f})].
\EEA
This is the standard solution of the maximal work extraction problem given in thermodynamics
\cite{landau,callen}.

Let us now adopt the above reasoning assuming that the overall system
consists of a large thermal bath B and a small system R.  Initially B
is in equilibrium at inverse temperature $\beta$, while R
is in an arbitrary state $\rho$. Thus the initial state is
\BEA
\Phi = \rho\otimes \rho_{B}(\beta),
\qquad \rho_{B}(\beta)=\frac{1}{Z_{\rm B}}\, e^{-\beta H_{\rm B}},
\EEA
where $H_{\rm B}$ is the bath Hamiltonian. 
The overall initial Hamiltonian is
\BEA
H= H_{\rm R}\otimes 1+1\otimes H_{\rm B},
\EEA
where $H_{\rm R}$ is the Hamiltonian of R.

We now employ the following facts:\\ 
(1.) The number of bath degrees of
freedom is much larger than that for the small system R.  \\
(2.) Since the
bath started from a passive equilibrium state, the extracted work is
expected to be of the same order as the energy of R.  Thus the extracted
work is much smaller than the bath energy.  \\
(3.) Since $\beta = {\cal O}(H_{\rm R})$, the difference between $\beta$ and
$\beta_{\rm f}$ is expected to be much smaller than $H_{\rm R}$.  \\
(4.)
The interaction between B and R occurs only during the work-extraction
process.  It is negligible both before and after the process. The
interaction may be sizable during the work-extraction, but the
corresponding energy costs for switching it on and off are already
included in the work. 

Let us now write (\ref{ga2}) as
\BEA
\label{bratva}
S[\rho]+S[\rho_B(\beta)]=
S[\rho(\beta)]+S[\rho_B(\beta_{\rm f})],
\EEA
where $\rho(\beta)\propto \exp[-\beta H_{\rm R}]$, and 
where the LHS follows from the fact that initially the bath and the small system
were in a factorized state.  The RHS follows from the Gibbs density
matrix (\ref{kalimantan}) of the final state, taking into account in the
final state R and B do not interact. 

In (\ref{bratva}) we neglected the difference between
$\beta$ and $\beta_{\rm f}$ in $S[\rho(\beta)]$.  For
the bath this difference should not be neglected, since this small
difference is multiplied by the large number of the bath degrees of
freedom. We write
\BEA
\label{ga3}
S[\rho_B(\beta_{\rm f})]= S_B(U_B+\delta U_B),
\EEA
where $U_B$ is the initial bath energy, and where $\delta U_B$ is
the change of the bath energy due to its interaction with the small
system. Using $\d S_B/\d U_B = \beta$ and expanding 
\BEA
S_B(U_B+\delta U_B) =S_B(U_B)+\beta\delta
U_B, 
\EEA
we get from (\ref{bratva}):
$\delta U_B=T(\,S[\rho]-S[\rho(\beta)]\,)$.
Putting this into (\ref{ga4}) we get
\BEA
W_{\rm th}
&=&{\rm tr}\, [H\rho]
-{\rm tr}\, [H\rho(\beta)]-T(\,
S[\rho]-S[\rho(\beta)]\,)~~~~~\\
&=& F_{\rm i}-F_{\rm f}.
\label{bekor}
\EEA
This is the difference $F_{\rm i}-F_{\rm f}$
between the free energies, provided the latter is defined as
\BEA
F={\rm tr}\, [H\rho]-TS[\rho].
\EEA

Note that $W_{\rm th}$ in (\ref{bekor}) can be written as
$W_{\rm th}=T\,S[\,\rho\, ||\, \rho (\beta)\,]$,
where $S[\,\rho\, ||\, \rho (\beta)\,]$ defined in (\ref{relative_entropy}), 
is the relative entropy between the initial state $\rho$ and the final
equilibrium state $\rho(\beta)$ of R.

\subsection{Reachability of the thermodynamical upper bound via a slow process.}
\label{app2}

One may hope that the thermodynamical bound (\ref{bekor}) could indeed
be reached by some realistic work extraction dynamics
\cite{landau,callen,abn,alicki}. This is because for a macroscopic
system (for the present case R+B) the conservation of entropy alone is expected
to characterize a thermally isolated process. More specifically, we
should demand from a physical realization of the bound (\ref{bekor})
that the work-sources act on the system and, at worst, at the system
bath-coupling, but not on the bath itself, since the latter is supposed
to be out of a direct control. 

A realization of the bound (\ref{bekor}) for restricted initial states
$\rho$ was outlined in \cite{alicki}. This realization takes a long time
and thus amounts to zero power of work-extraction.  A finite power
protocol of extracting (\ref{bekor}) from an arbitrary initial state
$\rho$ is given in \cite{aa}.  However, this protocol has another
serious drawback, namely one has to allow direct interactions between
the source of work and the thermal bath (uncontrollable degrees
of freedom). 

The result presented in \cite{alicki} focussed on the zero
Hamiltonian case $H_{\rm R}=0$. It is, however, possible to generalize this
result assuming that the initial state $\rho$ is diagonal in the
representation of $H_{\rm R}$. This restriction is essential, as we
discuss below. Thus we write $H_{\rm R}$ and $\rho$ as
\BEA
\label{kro100}
&&H=\diag [\eps_n,...,\eps_1], \quad \eps_n\geq ...\geq \eps_1=0, \\
\label{kro101}
&&\rho=\diag [r_n,...,r_1],
\EEA
where $\diag[a,..,b]$ is a diagonal matrix with entries 
$(a,...,b)$. For simplicity we assume that the eigenvalues 
$r_k$ are ordered as
\BEA
\label{mangal}
r_n\leq ...\leq r_1.
\EEA
This assumption is not essential. If it does not hold, one should supplements 
the step {\bf 1} below by the unitary transformation that orders properly
the spectrum of $\rho$. 

The work extraction consists of the following two steps:

{\bf 1.} 
One changes with time the level spacings from their initial values
$\{\eps_n,...,\eps_2 \}$ to final values $\{\eps'_n,...,\eps'_2 \}$:
\BEA
\label{nunu}
\label{kagan2}
\{\eps_n,...,\eps_2 \}\to
\{\eps'_n,...,\eps'_2 \}.
\EEA
The change occurs much {\it faster} than the relaxation time induced by the bath,
so that the interaction with the bath can be neglected during the change.
Since the corresponding time-dependent Hamiltonian commutes with the
initial density matrix for all times, the populations $\{r_n,...,r_1 \}$
remain constant during this process. The purpose of (\ref{kagan2}) is to
achieve the {\it local equilibrium} of the system at the bath
temperature $\beta$:
\begin{gather}
\label{kagan3}
r_n=\frac{e^{-\beta\eps'_n  }}{Z}, \quad...,\quad
r_2=\frac{e^{-\beta\eps'_2  }}{Z},\quad
r_1=\frac{1}{Z},
\end{gather}
where $Z\equiv \sum_{k=1}^n e^{-\beta\eps'_k  }$.
Together with (\ref{kro100}, \ref{mangal}), Eqs.~(\ref{kagan3}) define the new spacings 
$\{\eps'_n,...,\eps'_2 \}$:
\BEA
\label{kagan33}
\eps'_n=T_{\rm i}\ln\frac{r_1}{r_n},...,\quad
\eps'_n=T_{\rm i}\ln\frac{r_1}{r_2}.
\EEA 

The full work $W_1$ done during (\ref{kagan2}) 
is [initial average energy minus final average energy]
\BEA
W_1=\sum_{k=2}^n r_k [\eps_k -\eps'_k].
\label{arktur}
\EEA

{\bf 2.} 
The spacings $\eps'_n,...,\eps'_2$ are now slowly moved back to their
original values $\eps_n,...,\eps_2$. Here slow means that the
characteristic time of the variation is much larger than the relaxation
time of the system (determined by the coupling to the bath). It is at
this point that the work-extraction process is going to take a long
time. 

During this process the density matrix is
\BEA
\label{tartar}
&&\rho_2(t)=\frac{1}{Z(t)}\,\diag \left[
e^{-\beta\eps_n(t)  },\quad...\quad,
e^{-\beta\eps_2(t)  },
1 \right],~~~ \\
&&Z(t)\equiv \sum_{k=1}^n e^{-\beta\eps_k(t)  }, 
\label{khozuk}
\EEA
where at the initial stage of this second step $\eps_k(t_{\rm i})=\eps'_k$,
while at the final stage $\eps_k(t_{\rm f})=\eps_k$.

The work done during this process reads
\begin{gather}
\label{lord}
W_2=-\int_{t_{\rm i}}^{t_{\rm f}}\d s\, \rho_2(s)
\partial_s H(s)=
\int_{t_{\rm i}}^{t_{\rm f}}\d s\, \frac{\d \left[\,T\ln Z(s)   \,\right]}{\d s},
\end{gather}
where $Z(t)$ is defined in (\ref{khozuk}).
Working out (\ref{lord}) one gets that
$W_2$ is equal to the free energy difference:
\BEA
\label{blat1}
W_2&=&-\left[\, {\rm tr}(\,\rho(\beta\,) H)-T\,S[\,\rho(\beta)\,]\,\right]\\
&+& \sum_{k=2}^n r_k \eps'_k + T \sum_{k=1}^n r_k\ln r_k,
\label{blat2}
\EEA
where the RHS of (\ref{blat1}) is the minus final free energy with
$\rho(\beta)=\rho_2(t_{\rm f})$ being the final equilibrium state of R,
and where (\ref{blat2}) is the free energy at the end of the 
sudden change defined in (\ref{nunu}).
It is clear that the sum $W_1+W_2$ defined via (\ref{arktur},
\ref{blat1}, \ref{blat2}) is equal to the thermodynamic bound
(\ref{bekor}). 

Since the local equilibrium is related to the simultaneously diagonal
form for both $H$ and $\rho_0$, would there be initial coherences
[non-diagonal elements of $\rho_0$], there would be no unitary operation
that could bring the system to the local equilibrium with the bath at
the end of the first step. Thus the adopted restriction on the initial
state is essential. 


\begin{thebibliography}{99}

\bibitem{landau}L.D. Landau and E.M. Lifshitz, {\it Statistical
Physics, I}, Pergamon Press Oxford, 1978.


\bibitem{callen}
H.B. Callen, {\it Thermodynamics}, (John Wiley, NY, 1985). 

\bibitem{zotin}
A.I. Zotin and I. Lamprecht, J. Theor. Biol. {\bf 180}, 207 (1996).

I. Lamprecht and A.I. Zotin, eds.,
{\it Thermodynamics and Kinetics of Biological Processes} 
(Walter de Gruyter, Berlin, 1982).

\bibitem{geo}H.  Ozawa {\it et al.}, Rev. Geophys. {\bf 41}, 4 (2003).


\bibitem{60}
H.E.D. Scovil and E.O. Schultz-DuBois, Phys. Rev. Lett., {\bf 2},
262 (1959).

J.E. Geusic et al., Phys. Rev., {\bf 156}, 343 (1967).

\bibitem{review}
V.K. Konyukhov and A.M. Prokhorov, Uspekhi Fiz. Nauk, {\bf 119}, 541 (1976).



\bibitem{kosloff}
R. Kosloff, J. Chem. Phys. 80, 1625 (1984).


\bibitem{geva}
E. Geva and R. Kosloff, J. Chem. Phys. {\bf 104}, 7681 (1996).
T. Feldmann and R. Kosloff, Phys. Rev. E {\bf 70}, 046110
(2004).



\bibitem{var}

C.M. Bender, D.C. Brody and B.K. Meister, J. Phys. A {\bf 33}, 4427 (2000). 
S. Lloyd, Phys. Rev. A, {\bf 56} (1997) 3374.
J. He et al., Phys. Rev E, {\bf 65}, 036145 (2002).


\bibitem{jmod}A.E. Allahverdyan, R. Balian and Th.M. Nieuwenhuizen, 
J. Mod. Opt. {\bf 51}, 2703 (2004).

\bibitem{nori}H. T. Quan, Y. D. Wang, Y. Liu, C.P. Sun and
F. Nori, Phys. Rev. Lett. {\bf 97}, 180402 (2006).

\bibitem{nori_c}H. T. Quan, Y. D. Wang, Y. Liu, C.P. Sun and
F. Nori, quant-ph/0611275.

\bibitem{beretta}
G.P. Beretta, quant-ph/0703261.

\bibitem{boukobza} E. Boukobza and D. Tannor, 
Phys. Rev. A {\bf 74}, 063823 (2006); Phys. Rev. A {\bf 74}, 
063822 (2006).

\bibitem{linke}T.E. Humphrey, R. Newbury, R.P. Taylor and H.
Linke, Phys. Rev. Lett. {\bf 89}, 116801 (2002).

\bibitem{linke_talk}
T.E. Humphrey and H. Linke, Physica E, {\bf 29}, 390 (2005).


\bibitem{ki} T. D. Kieu, Phys. Rev. Lett. {\bf 93}, 140403 (2004);
Eur. Phys. J. D {\bf 39}, 115 (2006).

\bibitem{scully}
M.O. Scully, M.S. Zubairy, G.S. Agarwal, and H. Walther,
Science {\bf 299}, 862 (2003). 

M. Scully, Phys. Rev. Lett. {\bf 87} (2001) 220601; 
{\it ibid} {\bf 88} (2002) 050602.

\bibitem{1} A. E. Allahverdyan, R. Serral Gracia, and T. M.
Nieuwenhuizen, Phys. Rev. E {\bf 71}, 046106 (2005).

\bibitem{2}
M. J. Henrich, M. Michel and G. Mahler, 
Europhys. Lett., {\bf 76}, 1057 (2006).

M. J. Henrich, G. Mahler and M. Michel, cond-mat/0705.0075.

\bibitem{alicki}
R. Alicki, M. Horodecki, P. Horodecki and R. Horodecki,
Open Syst. Inf. Dyn. {\bf 11}, 205 (2004).

\bibitem{muru}K. Maruyama, F. Nori and V. Vedral, physics/07073400.


\bibitem{partovi}H.M. Partovi, Phys. Lett. A, {\bf 137}, 440 (1989).


\bibitem{delgado}A. Galindo and M.A. Martin-Delgado, Rev. Mod. Phys.,
{\bf 74}, 347, (2002).

\bibitem{abn}A.E. Allahverdyan, R. Balian and Th.M. Nieuwenhuizen, 
Europhys. Lett. {\bf 67}, 565 (2004).


\bibitem{alicki_jpa}
R. Alicki, J. Phys. A, {\bf 12}, L103 (1979).







\bibitem{CA}F. Curzon and B. Ahlborn, Am. J. Phys. {\bf 43}, 22 (1975).

\bibitem{leff}

H. Leff, Am. J. Phys. {\bf 55}, 602 (1987). 

P. T. Landsberg and H. Leff, J. Phys. A {\bf 22}, 4019 (1989).


\bibitem{CA_van} C. Van den Broek, Phys. Rev. Lett. {\bf 95}, 190602 (2005).



\bibitem{finite_time}
R.S. Berry, V.A. Kazakov, S. Sieniutycz, Z. Szwast, and
A.M. Tsvilin, {\it Thermodynamic Optimization of Finite--Time Processes} 
(John Wiley \& Sons, Chichester, 2000).





\bibitem{saka}H. Sakaguchi, J. Phys. Soc. Jpn. {\bf 67}, 709 (1998).

\bibitem{astumian}
I. Derenyi and R. D. Astumian, Phys. Rev. E {\bf 59}, R6219 (1999).

M. Asfaw and M. Bekele, Eur. Phys. J. B {\bf 38},
457, (2004); arXiv:cond-mat/0605233.


\bibitem{jauslin}N. Sangouard, X. Lacour, S. Guerin, and H.R. Jauslin, 
quant-ph/0505163.

\bibitem{ando} A.W. Marshall and I. Olkin, {\it Inequalities: Theory
of Majorization and its Applications}, (Academic Press, New York,
1979).

\bibitem{vaks}A B Brailovskii, V.L. Vaks and V.V. Mityugov, 
Physics Uspekhi, {\bf 166}, 795 (1996).

\bibitem{mito}V.V. Mityugov, Physics Uspekhi, {\bf 170}, 681 (2000).

\bibitem{ziman}V. Scarani, M. Ziman, P. Stelmachovic, N. Gisin, and V.
Buzek, Phys. Rev. Lett. {\bf 88}, 097905 (2002).

\bibitem{aa}A.E. Allahverdyan, unpublished notes. 

\end{thebibliography}
\end{document}